\pdfoutput=1
\PassOptionsToPackage{dvipsnames}{xcolor}

\documentclass{article}

\usepackage[preprint]{htmlcure_preprint}

\usepackage{booktabs}
\usepackage[T1]{fontenc}
\usepackage[utf8]{inputenc}
\usepackage{hyperref}
\usepackage{url}
\usepackage{xurl}
\usepackage{microtype}
\IfFileExists{inconsolata.sty}{\usepackage{inconsolata}}{}

\usepackage{graphicx}
\usepackage{amsmath}
\usepackage{amssymb}
\usepackage{bm}
\usepackage{subcaption}
\usepackage{multirow}
\usepackage{enumitem}
\usepackage{colortbl}
\usepackage{xspace}
\usepackage{listings}
\usepackage{placeins}
\usepackage{float}
\usepackage{wrapfig}
\usepackage{afterpage}

\usepackage[most, breakable, skins]{tcolorbox}
\usepackage{xcolor}
\usepackage{pifont}
\usepackage{tikz}
\newtcblisting{promptbox}[1]{
  enhanced,
  colback=white,
  colframe=black!35,
  colbacktitle=black!6,
  coltitle=black,
  title={#1},
  fonttitle=\bfseries\footnotesize,
  listing only,
  boxsep=1pt,
  left=3pt,
  right=3pt,
  top=3pt,
  bottom=3pt,
  before skip=4pt,
  after skip=6pt,
  listing options={
    basicstyle=\ttfamily\tiny,
    breaklines=true,
    columns=fullflexible,
    keepspaces=true,
    showstringspaces=false
  }
}

\usetikzlibrary{arrows.meta,positioning,calc,shapes.geometric,fit,backgrounds}
\IfFileExists{fontawesome5.sty}{\usepackage{fontawesome5}}{\newcommand{\faLock}{\ensuremath{\blacklozenge}}}

\definecolor{ababcol}{HTML}{F14738}
\definecolor{myhailuo2}{HTML}{F97669}
\definecolor{querycol}{HTML}{7964E8}
\definecolor{goldanswercol}{HTML}{FFB43B}
\definecolor{otherscol}{HTML}{FC5BCF}
\definecolor{myhailuo3light}{HTML}{FFA9FA}
\definecolor{myhailuo1dark}{HTML}{FC8900}
\definecolor{myhailuo2dark}{HTML}{F14738}
\definecolor{myhailuo3dark}{HTML}{D12AAA}
\definecolor{myhailuo4dark}{HTML}{4C4DC2}

\definecolor{myhailuo1}{HTML}{FFB43B}
\definecolor{myhailuo2}{HTML}{F97669}
\definecolor{myhailuo3}{HTML}{FC5BCF}
\definecolor{myhailuo4}{HTML}{7964E8}

\definecolor{myhailuo1light}{HTML}{FFD085}
\definecolor{myhailuo2light}{HTML}{FFA19F}
\definecolor{myhailuo3light}{HTML}{FFA9FA}
\definecolor{myhailuo4light}{HTML}{BDACFB}

\colorlet{myorange}{Orange!20}
\colorlet{mygreen}{LimeGreen!25}
\colorlet{myyellow}{Yellow!30}
\colorlet{myblue}{CornflowerBlue!25}
\colorlet{mybrown}{RawSienna!25}
\colorlet{mypurple}{Orchid!25}
\colorlet{myred}{Red!60}
\colorlet{myorangefull}{YellowOrange!60}
\colorlet{mybrownfull}{RawSienna!60}

\colorlet{myorangethick}{Orange!40}
\colorlet{mygreenthick}{LimeGreen!50}
\colorlet{myyellowthick}{Yellow!60}
\colorlet{mybluethick}{CornflowerBlue!50}

\tcbset{
    showcase/.style={
        fonttitle=\large,
        colback=white!20,  
        colframe=black,   
        coltitle=white,   
        boxrule=0.5mm,    
        arc=2mm,          
        outer arc=2mm,    
        left=1mm,         
        right=1mm,        
        top=1mm,          
        bottom=1mm,       
        width=\textwidth, 
        before skip=0.1pt,
        after skip=0.1pt,
    },
    context/.style={
        fontupper=\scriptsize,
        fonttitle=\large,
        colframe=querycol,     
        coltitle=white,   
        colback=white,    
        boxrule=0.3mm,    
        arc=2mm,          
        outer arc=2mm,    
        left=1mm,         
        right=1mm,        
        top=1mm,          
        bottom=1mm,       
        before skip=1pt,
        after skip=0.1pt, 
    },
    query/.style={
        fontupper=\scriptsize,
        fontlower=\scriptsize,
        colframe=querycol,     
        coltitle=white,   
        colback=white,    
        boxrule=0.1mm,    
        arc=2mm,          
        outer arc=2mm,    
        left=1mm,         
        right=1mm,        
        top=1mm,          
        bottom=1mm,       
        before skip=1pt,
        after skip=0.1pt,
    },
    abab/.style={
        fontupper=\scriptsize,
        fonttitle=,
        colframe=ababcol, 
        coltitle=white,   
        boxrule=0.5mm,    
        arc=2mm,          
        outer arc=2mm,    
        left=1mm,         
        right=1mm,        
        top=1mm,          
        bottom=1mm,       
        width=0.33\textwidth, 
        before skip=0.1pt,
        after skip=0.1pt, 
    },
    others/.style={
        fontupper=\scriptsize,
        colframe=myhailuo3light, 
        coltitle=white,
        boxrule=0.5mm,    
        arc=2mm,          
        outer arc=2mm,    
        left=1mm,         
        right=1mm,        
        top=1mm,          
        bottom=1mm,       
        width=0.33\textwidth, 
        before skip=0.1pt,
        after skip=0.1pt, 
    },
    goldanswer/.style={
        fontupper=\scriptsize,
        colframe=goldanswercol,     
        coltitle=white,   
        boxrule=0.5mm,    
        arc=2mm,          
        outer arc=2mm,    
        left=1mm,         
        right=1mm,        
        top=1mm,          
        bottom=1mm,       
        width=0.33\textwidth, 
        before skip=0.1pt,
        after skip=0.1pt, 
    },
}

\newcommand{\method}{\textsc{HTMLCure}\xspace}
\newcommand{\benchmark}{\textsc{HTMLBench}\xspace}

\definecolor{TopColor}{HTML}{E3F2FD}
\definecolor{MidColor}{HTML}{FFF3E0}
\definecolor{LowColor}{HTML}{FFEBEE}
\definecolor{HeaderColor}{RGB}{70, 130, 180}
\definecolor{ExcellentColor}{RGB}{232, 245, 233}
\definecolor{GoodColor}{RGB}{250,250,209}
\definecolor{FairColor}{RGB}{255, 235, 238}
\definecolor{OurColor}{RGB}{230, 240, 255}
\arrayrulecolor{black}

\title{HTMLCure: Turning Browser Experience into State Guided Repair for Interactive HTML}

\author{%
  Jiajun Wu\textsuperscript{1},
  Jian Yang\textsuperscript{1},
  Tuney Zheng\textsuperscript{2},
  Wei Zhang\textsuperscript{1} \\
  Haowen Wang\textsuperscript{2},
  Yihang Lou\textsuperscript{3},
  Xianglong Liu\textsuperscript{1} \\
  \textsuperscript{1}Beihang University \\
  \textsuperscript{2}IQuest Research \quad
  \textsuperscript{3}Peking University
}

\begin{document}
\maketitle
\raggedbottom

\begin{abstract}
LLMs can now produce full HTML pages, but many of those pages are only superficially correct: they render once, then fail under scroll, hover, click, resize, or gameplay. Evaluation from screenshots can miss these failures, and filtering discards many pages that are still repairable. We introduce \method, a browser experience framework that evaluates HTML after the system has interacted with it. The evaluator executes the page across viewports and interaction states, records deterministic browser evidence, and gives the VLM curated keyframes from the executed trajectory rather than isolated screenshots. The same state signal drives a closed loop repair engine: \method diagnoses the current page, chooses a state specific repair family, runs each candidate again, and exports quality cleared pages for SFT. On a 97K prompt corpus, this expands the directly usable seed into a candidate pool of 63{,}703 quality cleared pages, from which we construct the final refined SFT set of 40K pages. Under the same backbone and training recipe, HTMLCure-27B-Refined reaches \textbf{50.6} on \benchmark-400 with \textbf{45.2}\% deterministic test case pass, placing it in the same performance band as strong reference rows such as Kimi-K2.6 and GPT-5.4. On the released MiniAppBench validation split, it reaches \textbf{81.2} average, improving raw 27B SFT by 15.3 points and approaching the level of strong reference systems.
\end{abstract}

\begin{figure*}[h]
\vspace{-10pt}
\centering
\begin{minipage}{0.80\textwidth}
\centering
\includegraphics[width=\linewidth]{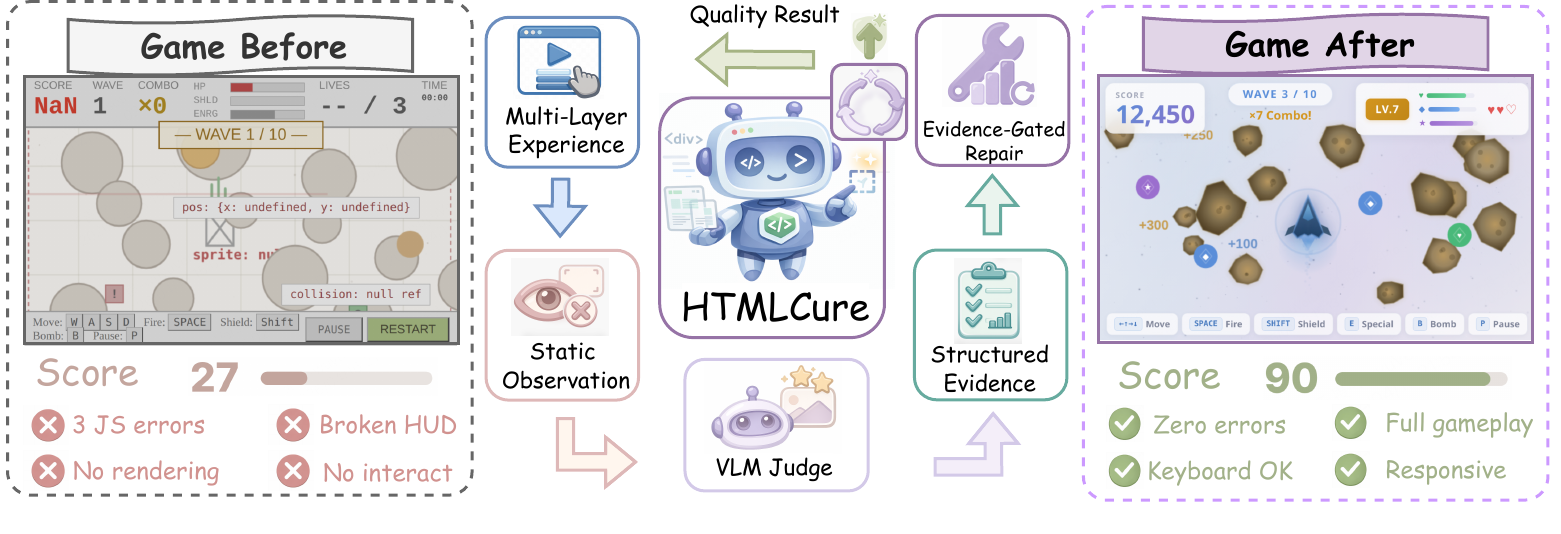}
\captionsetup{width=\linewidth}
\caption{Problem setting and full pipeline objective. \method evaluates interactive HTML through browser traces, routes repair by current page state, and exports only quality cleared pages into the refined SFT candidate pool.}
\label{fig:intro_overview}
\end{minipage}
\vspace{-8pt}
\end{figure*}

\section{Introduction}
\label{sec:intro}

LLMs can write frontend code at scale, but the generated HTML often looks acceptable in the first screenshot and then fails under use. In our 97K prompt corpus across six semantic task families, only about 15\% of Kimi-K2.5 pages are directly usable under the browser gate. The rest fail through inert controls, stalled game logic, broken responsive layouts, missing content, or export gate violations. Filtering removes these pages before testing whether they are recoverable. For data construction, this is a state estimation problem rather than only a scoring problem: a static screenshot or sparse visual trace can hide failures that appear only after hover, click, scroll, resize, or gameplay, leaving the repair loop with a weak basis for choosing among rewrite, local repair, and preservation.

\begin{wraptable}[9]{r}{0.47\textwidth}
\vspace{-17pt}
\centering
\scriptsize
\setlength{\tabcolsep}{2.2pt}
\renewcommand{\arraystretch}{1.06}
\setlength{\abovecaptionskip}{0pt}
\setlength{\belowcaptionskip}{4pt}
\caption{Repair action depends on current page state. ``Local fix'' groups diagnosis guided actions that do not rewrite the page. Success means $\Delta \geq 5$; catastrophe means $\Delta \leq -10$.}
\label{tab:intro_state_strategy}
\resizebox{\linewidth}{!}{%
\begin{tabular}{lccc}
\toprule
\textbf{State} & \textbf{Local fix} & \textbf{Rewrite} & \textbf{Policy} \\
\midrule
Low ($<40$)     & 92\%, +9.1         & 90\%, +14.2          & Rewrite \\
Mid (40--79)    & 95\%, +11.3        & 92\%, +8.3           & Diagnose \\
High ($\geq80$) & 0\%, $-0.2$        & 17\% cat., $-2.7$    & Refine \\
\bottomrule
\end{tabular}%
}
\vspace{-8pt}
\end{wraptable}

Table~\ref{tab:intro_state_strategy} shows the practical consequence: Low pages usually need broad replacement, Mid pages are better served by diagnosis guided local repair, and High pages should mostly be preserved.

We propose \method, a state aware repair and data construction framework built around this signal. \method first experiences an HTML page through the browser by rendering it, changing viewports, exercising controls, and probing deeper states when needed. The VLM judges curated keyframes from that trajectory. The novelty is not browser evaluation alone; prior benchmarks already use browser interaction. The contribution is to make the executed trace operational: the controller diagnoses the current page, selects a state specific repair family, runs each candidate again, and keeps candidates that pass the same quality gate. Figure~\ref{fig:framework} shows the full loop. \benchmark in \autoref{sec:benchmark} exposes the deterministic browser execution slice for reproducible comparison, while accepted repair traces become the refined SFT pool. Section~\ref{sec:exp} tests this data route under the same backbone and training recipe, with the released MiniAppBench validation split as an external check.

This leads to three contributions:

\textbf{(1) Browser experience as a repair state signal.} \method turns browser evidence into a structured page state that bounds VLM visual judgment and drives repair decisions.

\textbf{(2) State aware interactive repair.} The experienced state determines the repair family: rewrite weak pages, diagnose partially working pages, and preserve pages that are already strong. The controller runs candidates again and screens out repairs that regress under the quality gate.

\textbf{(3) Repair as data construction.} Accepted repairs expand the original High subset into a larger refined SFT corpus. Under the matched SFT recipe, the final 27B model reaches \textbf{50.6} on \benchmark-400 with \textbf{45.2}\% deterministic test case pass, putting it in the same range as Kimi-K2.6 and GPT-5.4, and reaches \textbf{81.2} average on the released MiniAppBench validation split, 15.3 points above raw 27B SFT.

\FloatBarrier

\section{Multi Dimensional Experiential Evaluation Pipeline}
\label{sec:eval}

\method evaluates an HTML page as an artifact that must run in a browser. The evaluator loads the page, changes the viewing context, exercises common interactions, and records the resulting evidence before any visual judgment is made. Figure~\ref{fig:framework} places this evaluator at the front of the pipeline. The output is not only a scalar score: it is an ordered experience trace that exposes the current page state and later drives repair routing.

\subsection{Design Philosophy}

\label{sec:eval_philosophy}

A screenshot based view reduces the page to \emph{single step visual observation},
\begin{equation}
\hat{y} \;=\; f\!\big(I(s_1), \ldots, I(s_k)\big),
\label{eq:single_step_eval}
\end{equation}
where $s_1,\ldots,s_k$ are sampled browser states, $I(s_i)$ is the rendered frame at state $s_i$, and $f$ is typically a visual judge. This view is useful for screenshot fidelity, but it leaves out failures that appear only after use. A page can render cleanly while click handlers fail, mobile layout breaks, animations stall, or console errors appear after a state change.

We instead evaluate the page by executing an action sequence before visual scoring. Let $h$ denote an HTML artifact and let $\pi = (a_1,\ldots,a_T)$ denote a sequence of viewing and interaction actions. The evaluator produces
\begin{equation}
\mathcal{E}(h,\pi) \;=\; \{e_t\}_{t=1}^{T}, \qquad e_t \;=\; (v_t, b_t),
\label{eq:experience}
\end{equation}
where $v_t$ records visual evidence such as rendered frames, viewport snapshots, and frame differences, and $b_t$ records behavioral evidence such as probe outcomes, DOM state, latency, and console state. The output is an ordered trace, not a small screenshot set. The visual component is judged from curated states after execution, while the deterministic probes keep the behavioral record explicit. The same trace supports repair because it localizes where the page failed, whether a later edit regressed, and what state the controller should act on in Section~\ref{sec:repair}.

\subsection{Four Layer Experience Protocol}
\label{sec:eval_protocol}

The protocol has four layers. Layers 1 and 2 cover ordinary use first: loading behavior, motion, layout under different device views, content below the fold, prominent controls, hover states, keyboard bindings, and scroll triggered behavior. These checks answer the question that matters before aesthetics: does the page render and respond under normal use?

When early evidence suggests deeper interactivity, Layer 3 runs a short task driven rollout so game states, counters, canvas changes, and multi step UI transitions remain visible to the evaluator. Layer 4 selects representative keyframes from the same run and attaches event context. The visual scorer sees the page through this executed trajectory rather than through unrelated still images.

\begin{figure*}[!t]
\vspace{-38pt}
\centering
\includegraphics[width=0.96\textwidth]{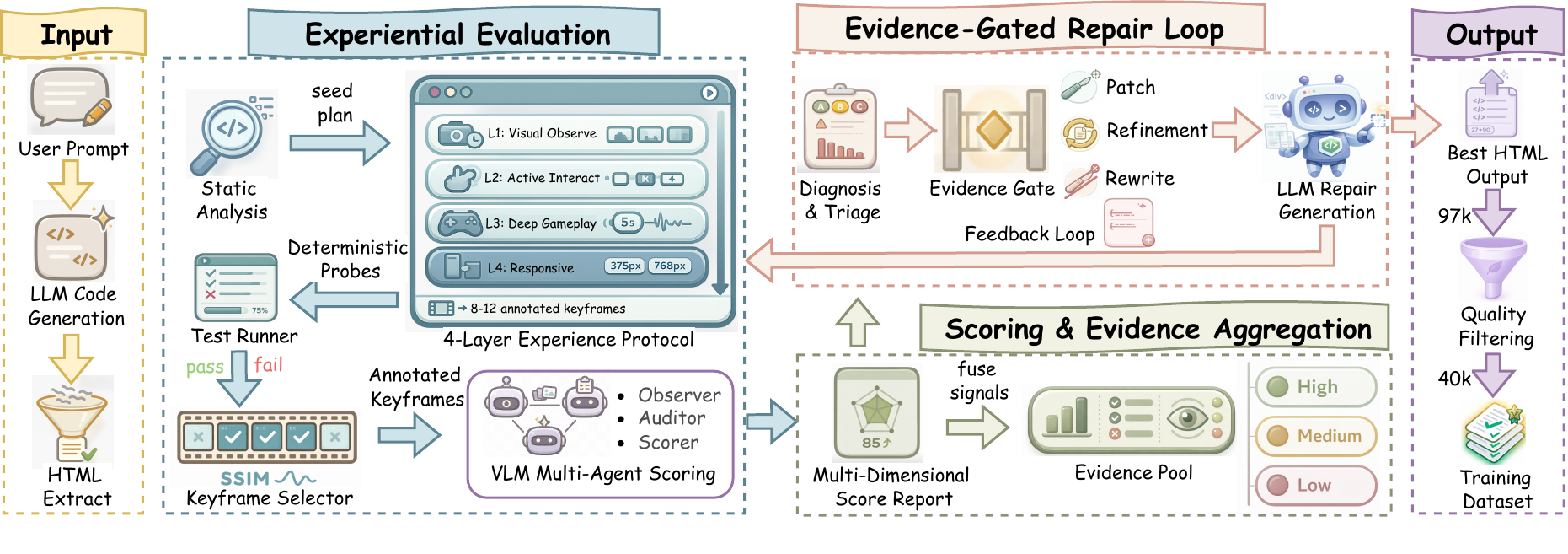}
\caption{\method pipeline. The evaluator gathers browser evidence and curated keyframes, the controller chooses a state specific repair family, and the export stage adds quality cleared pages to the refined SFT candidate pool before the final training export.}
\label{fig:framework}
\vspace{-13pt}
\end{figure*}

\subsection{Five Component Scoring}
\label{sec:eval_scoring}

We assign each page a five component score:
\begin{equation}
s_d(h)=g_d\!\big(\mathcal{E}(h,\pi)\big), \quad
S(h)=\sum_{d \in \mathcal{D}} s_d(h), \quad
\mathcal{D}=\{\text{rend}, \text{vis}, \text{func}, \text{inter}, \text{code}\}.
\label{eq:eval_score}
\end{equation}
Table~\ref{tab:scoring_dims} gives the public \benchmark scoring profile used in the final experiments. The mapping $g_d$ specifies how each dimension reads the trace. Coverage is recorded only as execution metadata; it is not a score dimension. Most points come from browser execution, deterministic test cases, or static checks; the VLM is reserved for visual design over curated states.

\begin{wraptable}[11]{r}{0.47\textwidth}
\vspace{-15pt}
\centering
\scriptsize
\setlength{\tabcolsep}{2.0pt}
\renewcommand{\arraystretch}{0.96}
\setlength{\abovecaptionskip}{0pt}
\setlength{\belowcaptionskip}{3pt}
\caption{\benchmark scoring profile. Coverage is metadata, not score.}
\label{tab:scoring_dims}
\resizebox{\linewidth}{!}{%
\begin{tabular}{lrrl}
\toprule
\textbf{Dimension} & \textbf{Inter.} & \textbf{Non-int.} & \textbf{Evidence} \\
\midrule
Rendering          & 10 & 10 & Browser health \\
\textbf{Visual Design} & \textbf{20} & \textbf{20} & \textbf{VLM keyframes} \\
Functionality      & 55 & 65 & Frozen TC pass \\
Interactivity      & 10 & 0  & Browser probes \\
Code Quality       & 5  & 5  & Static pass \\
\midrule
\textbf{Total}     & \textbf{100} & \textbf{100} & \textbf{Hybrid} \\
\bottomrule
\end{tabular}%
}
\vspace{-6pt}
\end{wraptable}

Equation~\eqref{eq:eval_score} makes this split explicit. Rendering, functionality, interactivity, and code quality are computed from browser execution or static analysis; visual design is the only component assigned through the VLM path. This path is bounded. An Analyst observes screenshots and probe evidence, while a separate Scorer receives no screenshots and assigns the visual component from the structured report plus objective metrics under deterministic decoding. Implementation guardrails reduce visual credit for horizontal overflow, missing viewport support, broken mobile layout, or absent styling. For noninteractive prompts, the interaction budget moves into functionality, so a static dashboard is not penalized for lacking controls it was never asked to provide. The scalar $S(h)$ sets the page state for repair, and the vector $(s_d(h))_{d \in \mathcal{D}}$ points to the dimension that limits the page. Leaderboard rows are therefore interpreted through the full evidence bundle: total score, deterministic test case pass rate, component scores, and the external MiniAppBench result.

\subsection{Cost Control and Tiered Gating}
\label{sec:eval_cost}

\method does not run every expensive probe on every page. A static pass removes malformed or trivial pages and collects code quality evidence. Layers 1--2 run on all remaining pages. Layer 3 runs only when earlier evidence indicates deeper interaction, and the visual scorer receives curated keyframes from Layer 4. Mid pages receive a diagnostic branch because routing decisions are hardest in that band. This keeps evaluation affordable while preserving state information needed by repair.

\section{State Aware Interactive Repair Engine}
\label{sec:repair}

The repair stage is where the experience trace becomes an intervention policy. A trace contains rendered states, interaction outcomes, console and probe evidence, device views, and curated keyframes. \method feeds this record to a repair controller before any generation call. The controller is not a generic regeneration prompt over source code or a screenshot. It first decides how much of the current page should be trusted, chooses a state specific edit family, and runs each candidate again before an accepted checkpoint can enter the refined pool. Figure~\ref{fig:framework} shows this controller branch between evaluation and export.

The first decision is how much of the current implementation to trust. A low scoring page often has no reliable local target and is better handled by replacement. A mid scoring page usually contains usable structure mixed with broken behavior, which makes diagnosis guided patching viable. A high scoring page is treated conservatively because broad rewriting can destroy working interaction. The VLM contributes contrastive before/after evidence, but the controller grounds each action in the full browser trace: probes, test failures, console state, device views, and curated visual evidence.

\subsection{Experienced State \texorpdfstring{$\to$}{->} Strategy Selection}
\label{sec:repair_principle}

Let $s = S(h)$ be the current score of page $h$, let $\mathbf{s}(h)=(s_d(h))_{d \in \mathcal{D}}$ be the component score vector, and let $D(h)$ be the structured diagnosis produced from the executed trace. The routing function $\sigma$ maps this experienced state to a repair family:
\begin{equation}
\sigma(D, s) \;=\;
\begin{cases}
\text{HolisticRewrite}, & s < 40, \\
\text{Rewrite or InteractionTargetedFix}(D), & 40 \leq s < 80, \\
\text{Preserve or TargetedRefine}(D), & s \geq 80.
\end{cases}
\label{eq:strategy}
\end{equation}

$D(h)$ is a structured controller input rather than a free form rationale. It contains failed test case identifiers, requirement statuses, component scores, render and console failures, responsive keys, button and form probe outcomes, gameplay or canvas evidence when present, keyframe annotations, and repair history. The controller used in our experiments is a fixed rule system over this record, not a learned policy. The 40 and 80 thresholds define the Low/Mid/High operating states used throughout the data funnel, and the 97 threshold is an export oriented early stop target. Within each band, observed failures bind the route to an admissible operator family.

The score bands turn the trust decision into a routing rule. Low pages are routed toward replacement because local edits rarely have a stable target. Mid pages keep access to both rewriting and diagnosis guided repair, since they often combine usable layout or logic with broken interactions. High pages are preservation cases: the default action is to keep the current checkpoint after rerunning the page, and small additive refinement is allowed only for a localized low risk defect. Thus, the trace is not passive evaluation output; it is the control state that determines which edits are allowed.

\subsection{Closed Loop Repair and Contrastive Feedback}
\label{sec:repair_loop}

Repair proceeds in short rounds. At round $t$, the system executes the current page $h_t$, forms the state summary $(S(h_t), \mathbf{s}(h_t), D(h_t))$, selects an admissible operator family, and prompts the LLM to produce up to two candidates $\mathcal{C}_t$. The small candidate budget keeps repair as a controlled edit process, not a search over many samples. The next state is chosen by
\begin{equation}
h_{t+1} \;=\; \arg\max_{h' \in \mathcal{C}_t \cup \{h_t\}} S_{\text{comp}}(h', h_t),
\label{eq:repair_select}
\end{equation}
where the inclusion of $h_t$ lets the controller keep the current page when all proposals score worse. The composite score $S_{\text{comp}}$ (\autoref{eq:composite}) penalizes regressions across dimensions instead of accepting a candidate only because one component improved. Candidate acceptance requires another browser run under the same protocol. Each round also uses \textbf{Contrastive Visual Feedback}: the VLM compares curated keyframes from $h_t$ and $h'$, then reports improved regions, regressions, and unchanged regions. The next prompt receives this visual diff together with deterministic browser evidence, so the LLM sees the concrete before/after trajectory that its previous edit produced.

\subsection{Constrained Strategy Space}
\label{sec:repair_strategies}

The admissible operator sets differ by state:
\begin{equation}
\begin{aligned}
\Omega_{\text{Low}} \;&=\; \{\text{HolisticRewrite}, \text{FeatureCompletion}, \text{GameRepair}\},\\
\Omega_{\text{Mid}} \;&=\; \{\text{Rewrite}, \text{BugFix}, \text{PlayabilityRepair},\\
&\qquad \text{InteractionTargetedFix}, \text{GameRepair}, \text{VisualEnrichment}\},\\
\Omega_{\text{High}} \;&=\; \{\text{NoOpVerify}, \text{VisualPolish}, \text{InteractionEnhance},\\
&\qquad \text{FunctionalityRefine}, \text{CodeCleanup}\}.
\end{aligned}
\label{eq:operator_sets}
\end{equation}
Within each set, $D(h)$ binds observed failures to specific operator families. Console or probe failures activate bug fixing; broken keyboard or mouse bindings activate playability repair; stable logic with weak presentation activates visual enrichment. Game repair is used when the trace shows that the interactive scaffold is worth preserving. The High state set is narrow because strong pages are more likely to be harmed by broad edits than helped by them; in this band, verification without editing is a valid controller output.

\subsection{Regression Aware Acceptance and Termination}
\label{sec:repair_antiregression}

Interactive repair can improve one dimension while damaging another. \method ranks candidates with a regression aware objective rather than raw total score:
\begin{equation}
S_{\text{comp}}(h', h) \;=\; S(h') \;-\; \sum_{d \in \mathcal{D}} w_d \cdot \max\!\big(0,\; s_d(h) - s_d(h')\big),
\label{eq:composite}
\end{equation}
where $h$ is the current page and $h'$ is a repaired candidate. Functionality and interactivity receive larger regression weights because failures in these dimensions are most visible during actual use. This objective is used in \autoref{eq:repair_select}, so an edit must survive another execution rather than win through a one sided gain in visual quality.

A trace stops when the page reaches a high target score ($s \geq 97$), when recent iterations show small or negative gains under the patience rule, or when the eight round budget is exhausted. Crossing 80 changes the admissible family from repair to refinement, but it does not stop the loop. High scoring pages stay under operators that preserve the working implementation.

\section{\benchmark: Deterministic Browser Executable Evaluation}
\label{sec:benchmark}

\begin{wraptable}[10]{r}{0.4\textwidth}
\vspace{-16pt}
\centering
\footnotesize
\setlength{\tabcolsep}{3.0pt}
\renewcommand{\arraystretch}{1.08}
\setlength{\abovecaptionskip}{0pt}
\setlength{\belowcaptionskip}{4pt}
{\captionsetup{font=scriptsize}
\caption{\benchmark task family composition.}
\label{tab:bench_categories}}
\resizebox{\linewidth}{!}{%
\begin{tabular}{lrrr}
\toprule
\textbf{Task family} & \textbf{Items} & \textbf{TCs} & \textbf{Subtypes} \\
\midrule
Apps \& Tools              & 105 & 1{,}688 & 18 \\
Content \& Marketing       & 110 & 1{,}660 & 16 \\
Data Visualization         & 35  & 588     & 7 \\
Games \& Simulations       & 55  & 682     & 9 \\
3D/WebGL Scenes            & 20  & 328     & 4 \\
Visual Art \& Animation    & 75  & 1{,}054 & 11 \\
\midrule
\textbf{Total}         & \textbf{400} & \textbf{6{,}000} & \textbf{65} \\
\bottomrule
\end{tabular}
}
\vspace{0.5pt}
\end{wraptable}

\benchmark is a standalone benchmark for single file interactive HTML generation. It is built to measure whether a generated page renders, satisfies prompt grounded requirements, responds to user actions, adapts across viewports, and maintains visual quality. The frozen release contains 400 item ids with category metadata, scored test cases, weights, and browser action sequences. Table~\ref{tab:bench_categories} gives the six user facing task families. The items cover 65 subtypes, 122 easy, 156 medium, and 122 hard prompts; 338 tasks require interaction because the benchmark is designed for pages that must respond under use, not only for static screenshot fidelity. Each scored test case is a deterministic browser program built from actions such as click, type, hover, key press, resize, screenshot change check, JavaScript assertion, and visibility check. Test cases are prompt grounded and avoid selectors, class names, frameworks, hidden source code assumptions, real credentials, payments, and private services that are specific to one implementation. A page's functionality score is the weighted pass rate over the frozen 6{,}000 test pool; coverage is kept only as execution metadata. The full leaderboard score remains hybrid because visual design uses the bounded VLM path from \autoref{sec:eval}, while TC pass reports the deterministic browser test slice. The benchmark files pass a strict schema validator and a benchmark quality audit for duplicate evidence, shallow visual checks, and cross template leftovers. An exact normalized prompt check finds no \benchmark prompt duplicated in the 91{,}484 traced repair records used for the SFT data analysis. Compared with agent based evaluation~\citep{zhang2026miniappbench}, \benchmark provides reproducible, multi dimensional HTML quality measurement with localized browser level failure records.
\begin{table}[t]
\vspace{-20pt}
  \centering
  \scriptsize
  \setlength{\tabcolsep}{2.2pt}
  \renewcommand{\arraystretch}{0.99}
  \newcommand{\resultgroup}[1]{%
      \arrayrulecolor{black!65}\specialrule{0.055em}{0pt}{0pt}%
      \rowcolor{HeaderColor}
      \multicolumn{13}{c}{\rule[-0.30ex]{0pt}{2.05ex}\smash{\textcolor{white}{\textbf{#1}}}}\\[-0.8pt]
      \arrayrulecolor{black!65}\specialrule{0.055em}{0pt}{0pt}%
      \arrayrulecolor{black}%
  }
  \caption{Main results on \benchmark and the released MiniAppBench validation split. MoE sizes report total and active parameters when disclosed; equal size SFT controls appear only in Table~\ref{tab:ablation_results}.}
  \label{tab:main_results}
  \resizebox{0.98\textwidth}{!}{%
    \begin{tabular}{ll rrrrrrr rrrr}
      \toprule
      \multirow{2}{*}{\textbf{Model}} &
      \multirow{2}{*}{\textbf{Size}} &
      \multicolumn{7}{c}{\textbf{\benchmark-400}} &
      \multicolumn{4}{c}{\textbf{MiniAppBench-Val}} \\
      \cmidrule(lr){3-9} \cmidrule(lr){10-13}
      & & \textbf{Score} & \textbf{TC Pass (\%)} &
      \textbf{Rend.} & \textbf{Vis.} & \textbf{TC/Func.} & \textbf{Inter.} & \textbf{Code} &
      \textbf{Intent} & \textbf{Static} & \textbf{Dyn.} & \textbf{Avg.} \\
      \resultgroup{Open source models}
      \rowcolor{GoodColor}
      Kimi-K2.6 & 1T${\mathrm{A}}$32B & \textbf{49.8} & \textbf{43.7} & 8.7 & 10.8 & \textbf{24.8} & 1.2 & \textbf{4.1} & 92.2 & 85.6 & 78.9 & 85.5 \\
      \rowcolor{GoodColor}
      Kimi-K2.5 & 1T${\mathrm{A}}$32B & 49.4 & 42.5 & 9.1 & 11.2 & 24.1 & 1.1 & 4.0 & 89.6 & 82.9 & 76.2 & 82.9 \\
      \rowcolor{GoodColor}
      GLM-5 & 744B${\mathrm{A}}$40B & 49.2 & 41.9 & 8.9 & 11.2 & 23.7 & 1.3 & 4.0 & 90.9 & 81.0 & 74.2 & 82.0 \\
      \rowcolor{GoodColor}
      DeepSeek-V3.2 & 685B${\mathrm{A}}$37B & 48.8 & 42.8 & 8.5 & 10.5 & 24.3 & \textbf{1.4} & \textbf{4.1} & 89.8 & 81.2 & 75.1 & 82.0 \\
      \rowcolor{GoodColor}
      GLM-5.1 & 754B${\mathrm{A}}$40B & 48.6 & 39.6 & 9.2 & \textbf{11.7} & 22.5 & 1.2 & 4.0 & \textbf{92.8} & \textbf{87.0} & \textbf{83.6} & \textbf{87.8} \\
      \rowcolor{GoodColor}
      GLM-4.7 & 358B${\mathrm{A}}$32B & 47.7 & 40.7 & 8.9 & 10.6 & 23.1 & 1.1 & \textbf{4.1} & 88.8 & 82.2 & 73.0 & 81.3 \\
      \rowcolor{GoodColor}
      Qwen3.6-27B & 27B & 47.1 & 38.1 & \textbf{9.3} & 11.0 & 21.6 & 1.1 & 4.0 & 91.7 & 84.5 & 76.8 & 84.3 \\
      \rowcolor{GoodColor}
      Qwen3.6-35B-A3B & 35B${\mathrm{A}}$3B & 46.3 & 37.4 & \textbf{9.3} & 10.8 & 21.2 & 1.0 & 4.0 & 90.0 & 83.1 & 72.2 & 81.8 \\
      \rowcolor{GoodColor}
      Qwen3.5-122B-A10B & 122B${\mathrm{A}}$10B & 45.6 & 38.0 & 9.0 & 10.0 & 21.5 & 1.0 & \textbf{4.1} & 87.2 & 80.7 & 77.8 & 81.9 \\
      \rowcolor{GoodColor}
      Qwen3.5-397B-A17B & 397B${\mathrm{A}}$17B & 45.6 & 37.3 & 9.2 & 10.1 & 21.1 & 1.1 & \textbf{4.1} & 84.4 & 77.3 & 61.7 & 74.5 \\
      \rowcolor{GoodColor}
      Qwen3.5-27B & 27B & 45.5 & 38.4 & 8.7 & 9.8 & 21.8 & 1.1 & 4.0 & 87.6 & 73.7 & 68.3 & 76.5 \\
      \rowcolor{GoodColor}
      Qwen3.5-35B-A3B & 35B${\mathrm{A}}$3B & 45.2 & 37.7 & 9.0 & 9.8 & 21.3 & 1.0 & \textbf{4.1} & 82.9 & 74.3 & 59.0 & 72.1 \\
      \rowcolor{GoodColor}
      DeepSeek-V4-Flash & 284B${\mathrm{A}}$13B & 44.5 & 35.7 & 8.5 & 10.7 & 20.3 & 1.1 & 3.9 & 90.1 & 82.0 & 83.1 & 85.1 \\
      \rowcolor{GoodColor}
      MiniMax-M2.5 & 229B${\mathrm{A}}$10B & 44.5 & 36.9 & 8.9 & 9.8 & 20.9 & 0.9 & 4.0 & 90.0 & 83.2 & 76.3 & 83.2 \\
      \rowcolor{GoodColor}
      Qwen3.5-9B & 9B & 42.9 & 35.1 & 8.9 & 9.1 & 19.9 & 0.9 & 4.0 & 73.3 & 63.2 & 30.7 & 55.7 \\
      \rowcolor{GoodColor}
      Qwen3.5-4B & 4B & 41.7 & 35.4 & 8.7 & 8.5 & 19.6 & 0.9 & 4.0 & 67.2 & 57.2 & 20.3 & 48.2 \\
      \resultgroup{Closed source models}
      \rowcolor{GoodColor}
      GPT-5.4 & \faLock & \textbf{49.2} & \textbf{42.7} & 8.8 & 11.0 & \textbf{24.2} & \textbf{1.0} & \textbf{4.2} & \textbf{90.8} & \textbf{86.5} & 85.6 & \textbf{87.7} \\
      \rowcolor{GoodColor}
      Claude-Opus-4.7 & \faLock & 47.8 & 39.8 & 9.2 & \textbf{11.1} & 22.6 & \textbf{1.0} & 3.9 & 89.7 & 83.3 & \textbf{85.9} & 86.3 \\
      \rowcolor{GoodColor}
      Claude-Opus-4.6 & \faLock & 47.5 & 39.3 & 9.2 & 11.0 & 22.3 & 0.9 & 4.1 & 87.3 & 81.8 & 81.9 & 83.7 \\
      \rowcolor{GoodColor}
      Claude-Sonnet-4.6 & \faLock & 46.6 & 37.2 & \textbf{9.4} & \textbf{11.1} & 21.1 & \textbf{1.0} & 3.9 & 87.9 & 83.4 & 80.8 & 84.0 \\
      \rowcolor{GoodColor}
      Gemini-3.1-Pro & \faLock & 45.4 & 37.0 & 8.9 & 10.5 & 21.0 & 0.9 & 4.0 & 84.3 & 78.8 & 78.3 & 80.5 \\
      \rowcolor{GoodColor}
      Claude-Opus-4.5-20251101 & \faLock & 43.9 & 38.5 & 8.8 & 8.9 & 21.8 & \textbf{1.0} & 3.3 & 88.0 & 82.2 & 77.8 & 82.7 \\
      \resultgroup{HTMLCure SFT models}
      \rowcolor{FairColor}
      HTMLCure-4B-Raw & 4B & 42.8 & 39.5 & 8.2 & 9.1 & 20.7 & 1.0 & 3.8 & 66.2 & 58.4 & 29.3 & 51.3 \\
      \rowcolor{FairColor}
      HTMLCure-4B-Filtered & 4B & 45.1 & 37.5 & 9.1 & 9.6 & 21.3 & 1.0 & 4.1 & 69.8 & 60.0 & 27.2 & 52.3 \\
      \rowcolor{OurColor}
      \textbf{HTMLCure-4B-Refined} & \textbf{4B} & \textbf{46.5} & 40.9 & 8.9 & 9.7 & 22.7 & \textbf{1.2} & 4.0 & 72.5 & 63.4 & 29.0 & 55.0 \\
      \rowcolor{FairColor}
      HTMLCure-9B-Raw & 9B & 43.0 & 41.4 & 8.0 & 8.9 & 21.3 & 1.0 & 3.8 & 72.1 & 59.3 & 20.3 & 50.6 \\
      \rowcolor{FairColor}
      HTMLCure-9B-Filtered & 9B & 45.8 & 40.3 & 8.7 & 9.6 & 22.4 & 1.1 & 4.0 & 73.7 & 63.9 & 27.1 & 54.9 \\
      \rowcolor{OurColor}
      \textbf{HTMLCure-9B-Refined} & \textbf{9B} & \textbf{48.5} & 42.1 & 9.2 & 10.2 & 23.9 & 1.1 & \textbf{4.1} & 76.5 & 65.0 & 28.1 & 56.5 \\
      \rowcolor{FairColor}
      HTMLCure-27B-Raw & 27B & 45.8 & 42.6 & 8.4 & 10.0 & 22.7 & 0.9 & 3.8 & 76.0 & 64.8 & 56.7 & 65.9 \\
      \rowcolor{FairColor}
      HTMLCure-27B-Filtered & 27B & 48.1 & 40.8 & \textbf{9.3} & 10.5 & 23.1 & 1.1 & \textbf{4.1} & 78.6 & 69.7 & 61.2 & 69.9 \\
      \rowcolor{OurColor}
      \textbf{HTMLCure-27B-Refined} & \textbf{27B} & \textbf{50.6} & \textbf{45.2} & 9.1 & \textbf{10.7} & \textbf{25.6} & 1.1 & \textbf{4.1} & \textbf{87.8} & \textbf{81.1} & \textbf{74.6} & \textbf{81.2} \\
      \bottomrule
    \end{tabular}%
  }%
  \vspace{-10pt}
\end{table}

\section{Experiments}
\label{sec:exp}

\subsection{Experimental Setup}
\label{sec:exp_data}

\begin{wraptable}[8]{r}{0.56\textwidth}
\vspace{-16pt}
\centering
\footnotesize
\setlength{\tabcolsep}{3.0pt}
\renewcommand{\arraystretch}{1.08}
\setlength{\abovecaptionskip}{0pt}
\setlength{\belowcaptionskip}{4pt}
\caption{Refined pool.}
\label{tab:data_funnel}
\resizebox{\linewidth}{!}{%
\begin{tabular}{lrrrrr}
\toprule
\textbf{State} & \textbf{Traces} & \textbf{Reject} & \textbf{Partial} & \textbf{Export} & \textbf{Share} \\
\midrule
Low ($<40$)      & 15{,}686 & 5 (0.0\%)       & 7{,}205 (45.9\%)  & 8{,}476 (54.0\%)  & 13.3\% \\
Mid (40--79)     & 62{,}012 & 10 (0.0\%)      & 11{,}121 (17.9\%) & 50{,}881 (82.1\%) & 79.9\% \\
High ($\geq80$)  & 13{,}786 & 9{,}440 (68.5\%)& 0 (0.0\%)          & 4{,}346 (31.5\%)  & 6.8\% \\
\midrule
\multicolumn{2}{l}{Filter baseline} & \multicolumn{4}{r}{13{,}786 High pages} \\
\multicolumn{2}{l}{Refined pool} & \multicolumn{4}{r}{63{,}703 pages (+49{,}917); 40K train export} \\
\bottomrule
\end{tabular}
}
\vspace{-1pt}
\end{wraptable}

\begin{figure}[t]
\vspace{-15pt}
\centering
\includegraphics[width=0.96\textwidth]{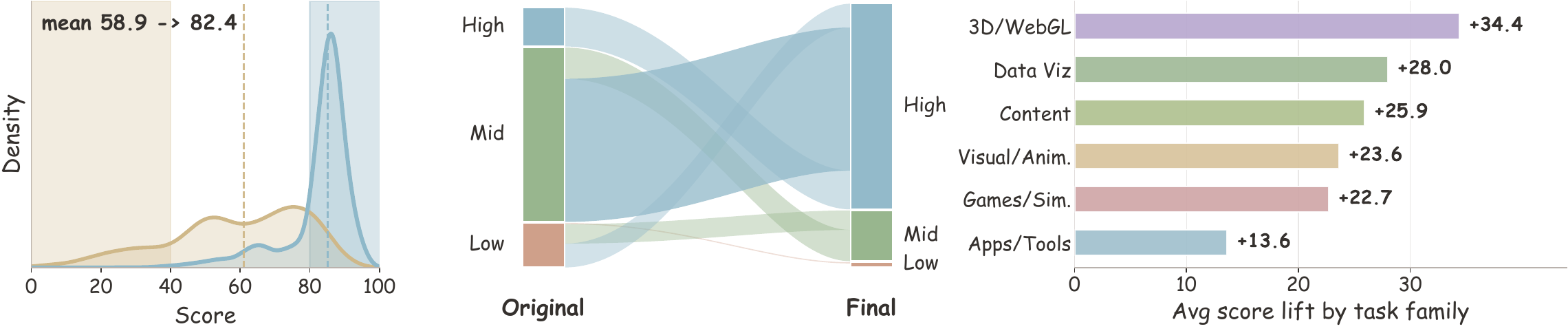}
\caption{Corpus level repair effect. Panel A compares the score density before and after repair, Panel B shows how traces migrate across the Low/Mid/High state bands, and Panel C reports mean lift by semantic task family. Taken together, the three views show a broad upward redistribution rather than a thin threshold effect.}
\label{fig:score_migration}
\vspace{-10pt}
\end{figure}

The experiments test whether repair from browser experience creates better supervision for SFT, rather than only higher page scores after repair. We use one browser crawled prompt corpus and construct three SFT routes from it. \textbf{Raw} keeps valid generated pages, \textbf{Filtered} keeps the original High subset, and \textbf{Refined} uses the final 40K export accepted after browser experienced repair. The routes use the same prompt response format and are evaluated with the same \benchmark runner and the released MiniAppBench validation split; the changing factor is how the HTML supervision is constructed. Table~\ref{tab:data_funnel} gives the refined pool funnel, while Appendix~\ref{app:sft_data} gives the extraction counts, state definitions, MiniAppBench protocol, and training configuration. The funnel rules out a simple filtering story: most exported pages come from repaired Low and Mid states, while High pages mainly serve as a preservation check. \autoref{fig:score_migration} gives the corresponding distributional view, showing that repair shifts the corpus upward rather than only selecting already good pages.

\FloatBarrier

\subsection{Main Results}
\label{sec:exp_main}

Table~\ref{tab:main_results} is the main downstream test. We fine tune Qwen3.5 4B, 9B, and 27B checkpoints~\citep{qwen35} with the same LlamaFactory recipe and prompt response format; Raw, Filtered, and Refined differ only in the source of HTML supervision, with Refined using the final 40K export cleared by the repair gate. On \benchmark-400, Refined leads the Raw/Filtered/Refined comparison at every scale, and HTMLCure-27B-Refined reaches 50.6 with 45.2\% deterministic test case pass. These two numbers should be read together: the total score includes the bounded visual component, while TC pass is the deterministic browser slice. On the released MiniAppBench validation split~\citep{zhang2026miniappbench}, the same model reaches 81.2 average, 15.3 points above raw 27B SFT. The main result is therefore the controlled route comparison: repair from browser experience turns weak and partial pages into verified supervision that improves the same backbone under the same training recipe. The reference rows provide model family context, not the causal evidence for the method.

\FloatBarrier

\subsection{Ablation Study}
\label{sec:exp_ablation}

\begin{wraptable}[10]{r}{0.50\textwidth}
\vspace{-15pt}
\centering
\scriptsize
\setlength{\tabcolsep}{3.2pt}
\renewcommand{\arraystretch}{0.96}
\setlength{\abovecaptionskip}{0pt}
\setlength{\belowcaptionskip}{2pt}
\caption{Equal size controlled ablation with 12{,}392 examples per route.}
\label{tab:ablation_results}
\resizebox{\linewidth}{!}{%
\begin{tabular}{lccccc}
\toprule
\textbf{Run} & \textbf{Scale} & \textbf{Data} & \textbf{Score} & \textbf{TC Pass (\%)} & \textbf{MiniApp} \\
\midrule
$A1$ & 9B  & Raw      & 41.9 & 37.8 & 53.7 \\
$A2$ & 9B  & Filtered & 45.1 & 37.0 & \textbf{55.2} \\
$A3$ & 9B  & Refined  & \textbf{47.5} & \textbf{40.9} & 55.1 \\
$A4$ & 27B & Raw      & 45.0 & 41.2 & 62.5 \\
$A5$ & 27B & Filtered & 48.2 & 41.0 & 71.8 \\
$A6$ & 27B & Refined  & \textbf{49.9} & \textbf{43.7} & \textbf{75.1} \\
\bottomrule
\end{tabular}
}
\vspace{-8pt}
\end{wraptable}

We next separate the amount of supervision from the kind of supervision produced by repair. Raw, Filtered, and Refined are matched to the same training count and retrained with the same recipe at 9B and 27B. Table~\ref{tab:ablation_results} reports this matrix; $A1$--$A6$ are only compact run identifiers. The matched comparison keeps the data count fixed and leaves the repair route as the changing factor. At 9B, Refined leads the matched routes on \benchmark score and pass rate, while MiniAppBench remains essentially tied with Filtered. At 27B, Refined leads the matched routes on both benchmarks under the same training count. This shows that browser experienced traces carry stronger supervision, not only more examples: pages interacted with, repaired under state specific control, and verified again still train the higher scoring model at matched corpus size.

\section{Analysis}
\label{sec:analysis}

\subsection{State Controls Repair Choice}
\label{sec:analysis_state}

We analyze 592{,}830 repair attempts over 91{,}484 pages to isolate how browser experience guides routing. A page with no stable structure, a page with broken controls, and a page that already passes most checks expose different repair risks. \autoref{tab:repair_mechanism} shows the consequence: the useful action family changes with the state observed by the browser trace.

\begin{table}[H]
\vspace{-10pt}
\centering
\scriptsize
\renewcommand{\arraystretch}{1.14}
\setlength{\abovecaptionskip}{0pt}
\setlength{\belowcaptionskip}{5pt}
\caption{State aware repair utility. ``Best route'' reports the highest lift action family in each state before repair; the rewrite columns show what happens when rewrite is used outside its useful regime.}
\label{tab:repair_mechanism}
\resizebox{0.98\textwidth}{!}{%
\begin{tabular}{llrrrrrrl}
\toprule
\textbf{State} & \textbf{Best route} & \textbf{$\Delta$} & \textbf{Success} & \textbf{Cat.} & \textbf{Rewrite $\Delta$} & \textbf{Rewrite cat.} & \textbf{$n$} & \textbf{Policy} \\
\midrule
Low ($<40$)  & Holistic rewrite & \textbf{+14.2} & 90.0\% & 0.6\%  & +14.2 & 0.6\%  & 35{,}822 & Rewrite weak pages \\
Mid (40--79) & Interaction fix   & \textbf{+11.3} & \textbf{95.4\%} & 0.1\% & +8.3  & 0.4\% & 324{,}880 & Diagnose then repair \\
High ($\geq80$) & Targeted fix & -0.2           & 0.0\%          & 0.0\% & -2.7 & 17.0\% & 232{,}128 & Conservative refine only \\
\bottomrule
\end{tabular}%
}
\vspace{-10pt}
\end{table}

Low pages are usually too incomplete for local patches, so broad rewriting has room to help. Mid pages already contain usable structure, and interaction repair gives a larger and safer local correction than rewriting the whole page. High pages are the preservation regime: their main risk is regression, not missing recovery. This is the first reason the controller cannot be reduced to a single generic repair prompt.

\begin{figure}[t]
\vspace{-15pt}
\centering
\includegraphics[width=0.96\textwidth]{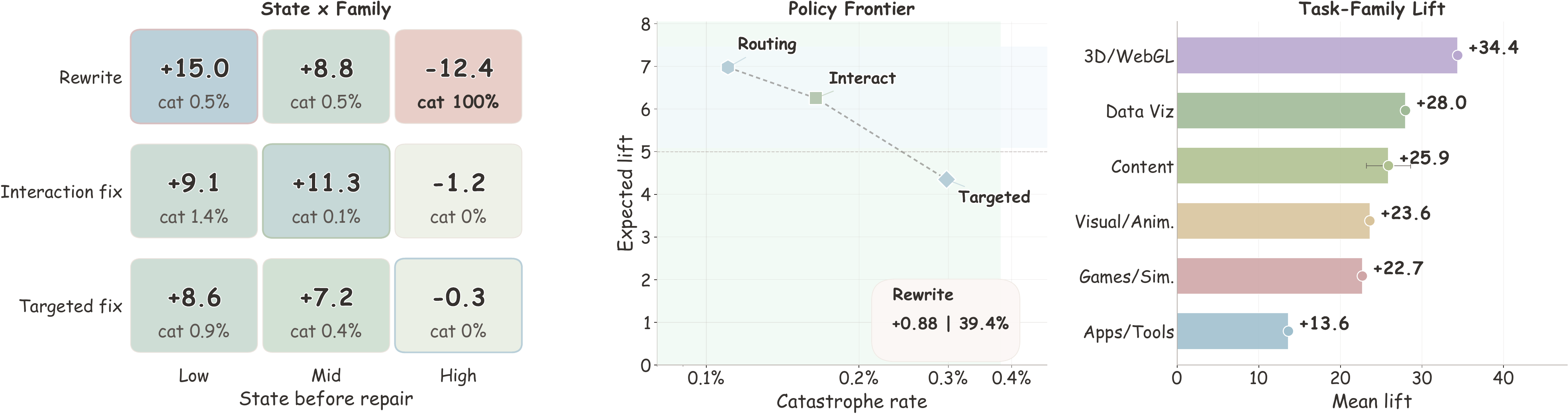}
\caption{Policy diagnostic for state aware repair. The figure compares action utility, fixed policies, and state aware routing across page states and task families.}
\label{fig:policy_diagnostic}
\vspace{-10pt}
\end{figure}
\noindent \autoref{fig:policy_diagnostic} turns the operator table into a policy diagnostic. Fixed targeted repair is safe but leaves many weak pages unrecovered. Fixed rewrite captures weak page recovery but pays for it on pages that should not be rewritten. State aware routing keeps the recovery side of rewrite while avoiding the high risk region for already working pages. Panel C is a useful sanity check rather than a separate claim: the lift varies by task family, but the preferred routing pattern remains stable enough to support one state conditioned controller.

\FloatBarrier

\subsection{Interactive Value and Checkpoint Retention}
\label{sec:analysis_rounds}

\begin{figure}[H]
\centering
\includegraphics[width=0.96\textwidth]{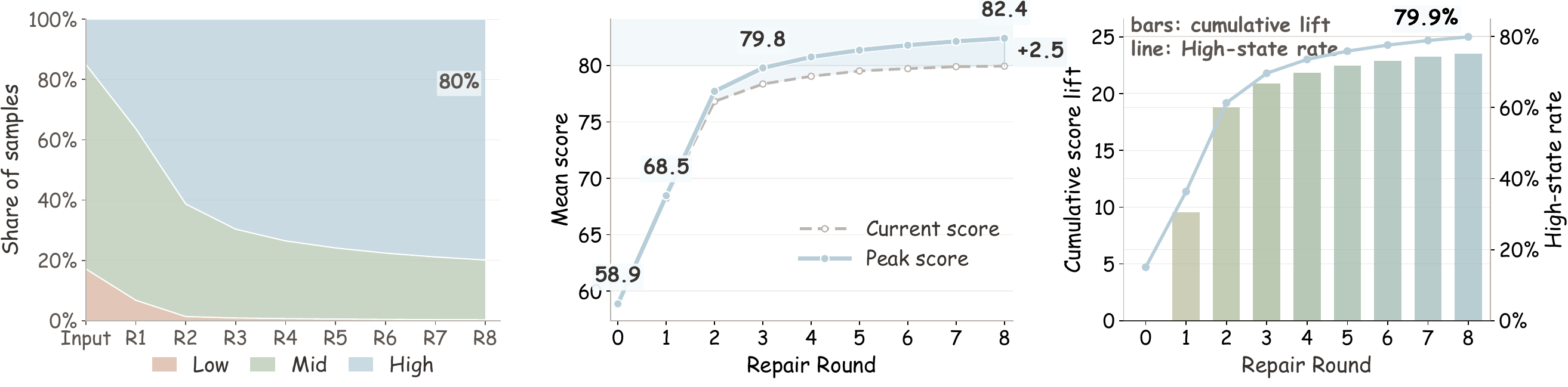}
\caption{Iteration level repair value. The figure shows how retained checkpoints improve over rounds, why best checkpoint export matters, and where repair gains saturate.}
\label{fig:repair_value_curve}
\end{figure}

State choice is only part of the mechanism. The repair loop also has to decide which candidate to keep. \autoref{fig:repair_value_curve} shows why \method exports the best verified checkpoint rather than the last candidate produced by the editor.

\noindent Panel A shows the population moving from Low and Mid toward High once the controller finds a useful path. Panel B is the main export diagnostic: after the early rounds, the current candidate can lag behind the retained checkpoint, so a "take the final edit" rule would discard verified gains. Panel C explains the stopping behavior. Most usable lift arrives early, but weaker pages still need a short tail of search. This keeps the claim calibrated: the value is not that every edit helps, but that browser re-execution lets the system reject regressions and export only cleared checkpoints.

\begin{figure}[t]
\centering
\vspace{-20pt}
\includegraphics[width=0.96\textwidth]{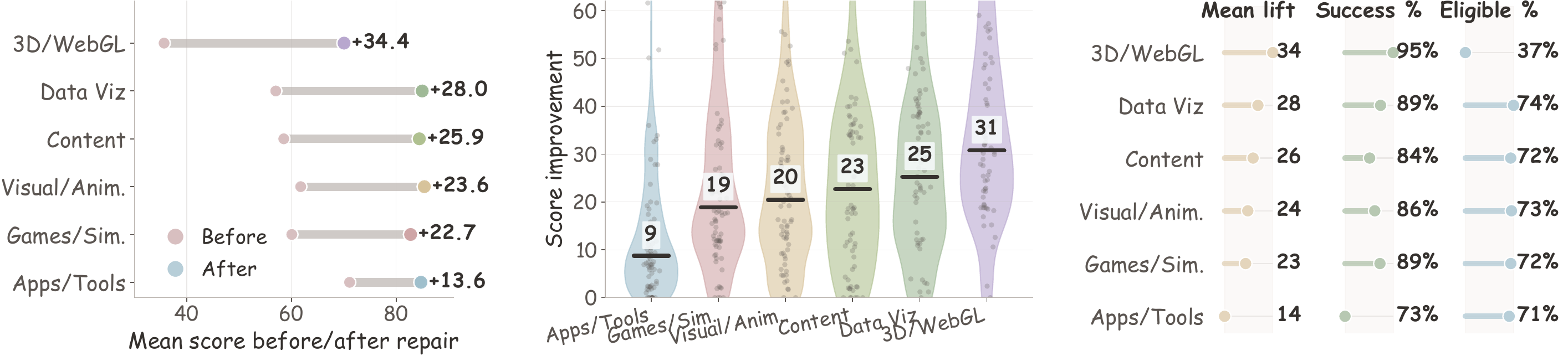}
\caption{Repair scope across semantic task families. The figure summarizes score gains, lift distributions, and final SFT eligibility by task family.}
\vspace{-10pt}
\label{fig:dataset_effectiveness}
\end{figure}
\noindent \autoref{fig:dataset_effectiveness} is a scope check. The task families fail in different ways, so the useful diagnostic is not a single mean lift. The figure shows that repair is not confined to one family, while also making clear that higher score does not automatically imply export. The score band accounting behind this view is defined in \autoref{tab:appendix_state_transition}; the SFT export gate remains the one reported in \autoref{tab:data_funnel}.

Together with the data funnel in \autoref{tab:data_funnel} and the corpus shift in \autoref{fig:score_migration}, these diagnostics give a compact mechanism story. Browser experience identifies the page state, state determines the repair family, and repeated execution protects the best recovered candidate. The main evidence for the method is the coupling between state aware routing, checkpointed export, and the downstream SFT gains in Tables~\ref{tab:main_results} and~\ref{tab:ablation_results}; the family level patterns serve as coverage checks rather than a separate causal claim.

\section{Related Work}
\label{sec:related}

\textbf{Web Agents and Browser Interaction.}
Language agent work studies how models interleave reasoning, actions, and feedback in external environments. ReAct introduces reasoning action traces, WebShop evaluates agents in a shopping website environment, and Reflexion uses trial feedback for later attempts~\citep{yao2023react,yao2022webshop,shinn2023reflexion}. Browser oriented benchmarks extend this line to realistic websites and multimodal browsing: Mind2Web, WebArena, and WebVoyager ask agents to complete tasks through web pages rather than answer from text alone~\citep{deng2023mind2web,zhou2024webarena,he2024webvoyager}. \method also treats the browser as an evidence source, but for a different objective. The trace is not the final trajectory of a web agent; it is the diagnostic state for routing HTML repair and selecting pages for supervised training.

\textbf{HTML Generation and Frontend Evaluation.}
Code intelligence has moved from foundation models toward agents and applications, as summarized in recent surveys and code model reports~\citep{yang2025codefoundationmodelsagents,yang2026iquestcoderv1technicalreport}. Frontend benchmarks measure complementary parts of HTML quality. Design2Code~\citep{si2025design2code} and Image2Struct~\citep{roberts2024image2struct} emphasize visual or structural fidelity, while MiniAppBench, WebGen-Bench, Vibe Code Bench, FullFront, Vision2Web, and WebCoderBench move toward generated interactive applications~\citep{zhang2026miniappbench,lu2025webgenbench,tran2026vibecodebench,sun2025fullfront,he2026vision2web,liu2026webcoderbench}. These benchmarks show why static screenshots or code similarity are insufficient for interactive pages. \method does not claim browser evaluation alone as the novelty. It uses the executed trace as a control signal: deterministic probes score most dimensions, the VLM judges bounded visual states from the trace, and the resulting state routes repair and export.

\textbf{Feedback Driven Repair and Synthetic Supervision.}
Self-Refine, Self-Debugging, CodeRL, and self-repair studies revise outputs from feedback or execution signals~\citep{madaan2023selfrefine,chen2023selfdebugging,le2022coderl,olausson2024selfrepair}; DesignBench and VIBEPASS study frontend repair and diagnosis in related settings~\citep{xiao2025designbench,bansal2026vibepass}. Separately, selected synthetic data can improve smaller models~\citep{gunasekar2023textbooks}, and web agent pipelines such as Explorer use exploration to synthesize or filter trajectories~\citep{pahuja2025explorer}. \method connects these directions for interactive HTML. A page may render yet fail under hover, canvas updates, button state changes, or mobile layout checks. Rather than only keeping or discarding it, browser experience classifies the current state, restricts the repair family, re-executes each candidate, and exports the best verified checkpoint as SFT supervision.

\section{Conclusion}
\label{sec:conclusion}

We introduce \method, a browser experience framework for evaluating and repairing LLM generated interactive HTML. The central idea is that execution should not be used only after generation as a pass or fail judge. Once a page has been rendered, interacted with, and viewed across states, the browser trace becomes a repair state signal: it separates missing structure from broken interaction and low risk refinement, selects the corresponding repair family, re-executes candidates, and exports the best verified checkpoint as SFT supervision. This turns repair into data construction rather than polishing after the fact. In our experiments, the refined pool trains a 27B model that reaches 50.6 on \benchmark-400 and 81.2 on the released MiniAppBench validation split, comparable to strong reference systems while preserving the controlled Raw/Filtered/Refined gains. The remaining limits are concrete. The VLM still judges the bounded visual design component, browser probes cannot cover every possible interaction, and complex games expose coupled timing and state failures. A next step is to learn routing and stopping from the repair traces themselves, while extending this approach beyond single file HTML.

\bibliography{custom}
\bibliographystyle{plain}

\clearpage
\appendix

\section{Limitations and Release Safeguards}
\label{sec:limitations}

The main evidence is the same backbone route comparison: Raw, Filtered, and Refined use the same base checkpoints, training recipe, and evaluation runners, so the changing factor is how the HTML supervision is constructed. This is the central axis of \method. The browser does not only score a page; it creates the state signal that decides whether the system should rewrite, diagnose and patch, or preserve the current implementation. The wider leaderboard provides context across model families, while the SFT block and the equal size controls isolate the repair data effect under matched training conditions. \benchmark provides a frozen, comprehensive evaluation suite for interactive HTML quality, and MiniAppBench supplies an external validation split. The visual design component is bounded, scored from structured evidence, and reported together with deterministic test case pass rate and MiniAppBench, so the main conclusion is tied to executed behavior rather than visual judgment alone.

The release is organized around reproducible artifacts. The artifact includes item JSONL files, frozen test cases, browser runner, scoring code, validation commands, configuration templates, aggregate trace tables, SFT manifests, and scripts for reconstruction. These files expose the benchmark, the execution based scoring path, the repair data funnel, and the aggregate evidence used by the figures and tables. For responsible release, benchmark prompts and generated examples avoid real credentials, payments, private services, and personal data; safety sensitive prompts are treated as UI only demos with synthetic placeholders. The main dual use risk is misuse of generated HTML for deceptive login, payment, or dashboard interfaces, so release documentation includes usage restrictions, filters for credential collection patterns, and license notices for Qwen, LlamaFactory, MiniAppBench, and provider generated outputs.

\section{Data and Code Availability}
\label{sec:data_code}

We provide the HTMLCure code and release materials at \url{https://github.com/wuyuVerse/HTMLCure}. The release contains the HTMLCure evaluation and repair code, the HTMLBench-400 benchmark files, the frozen 6{,}000 test case selection file, browser runner and scoring utilities, example configuration files, benchmark audit scripts, smoke tests, documentation for the evaluation protocol and architecture, aggregate trace tables, SFT manifests, provenance notes, reconstruction scripts, and an MIT license for the released HTMLCure code. For SFT data, the public release exposes route manifests and reconstruction scripts; release of the full generated and refined HTML corpus is handled under the applicable source and provider redistribution terms. External benchmarks, model APIs, Qwen checkpoints, MiniAppBench, LlamaFactory, and provider generated outputs remain governed by their original licenses and provider terms.

\section{Supplementary Analysis}

The main paper reports the aggregate effect of repair and moves the iteration level value curve into Section~\ref{sec:analysis_rounds}. This appendix keeps the lower level diagnostics: whether one strong repair prompt would be enough, and whether a fixed stopping rule would be safe. Both checks use the same browser traces that determine export into the refined SFT pool, so the evidence remains tied to observed rendering, interaction behavior, and regression risk.

\subsection{State Transition Accounting}

The state transition view is a score band diagnostic, not an export table. Rows in \autoref{tab:appendix_state_transition} use the original evaluator score before repair, and columns use the final retained score band after repair. Entries are percentages within each original state, with counts in parentheses. This table therefore asks whether repair moves pages across score bands under the browser evaluator. It does not say that every page in a final High band is exported into SFT; export also requires the quality gate and sampling step summarized in \autoref{tab:data_funnel}. This distinction matters most for original High pages: remaining in the High band is a preservation check, while Table~\ref{tab:data_funnel} reports how many of those pages are actually included in the refined pool.

\begin{table}[H]
\centering
\scriptsize
\renewcommand{\arraystretch}{1.10}
\setlength{\tabcolsep}{4pt}
\setlength{\abovecaptionskip}{0pt}
\setlength{\belowcaptionskip}{4pt}
\caption{State transition before and after repair. Rows are original score bands; columns are final retained score bands. Entries are within row percentages with counts.}
\label{tab:appendix_state_transition}
\resizebox{0.62\textwidth}{!}{%
\begin{tabular}{lccc}
\toprule
\textbf{Orig.} & \textbf{Low} & \textbf{Mid} & \textbf{High} \\
\midrule
Low  & 1.8\% (278)    & 44.2\% (6{,}932)  & 54.0\% (8{,}476) \\
Mid  & 0.0\% (0)      & 17.9\% (11{,}129) & 82.1\% (50{,}883) \\
High & 0.0\% (0)      & 0.0\% (0)         & 100.0\% (13{,}786) \\
\bottomrule
\end{tabular}%
}
\end{table}

\subsection{Strategy Profiles}

\begin{figure}[H]
\centering
\includegraphics[width=0.94\textwidth]{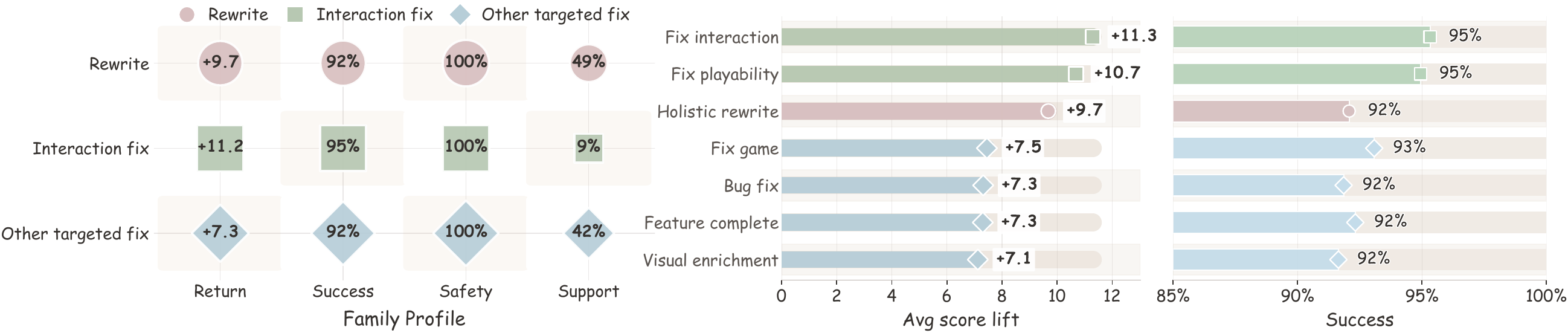}
\caption{Strategy profile for repairable pages. Panel A compares action families over return, success, safety, and support. Panels B and C rank concrete strategies by lift and success, showing that the useful action depends on the page state exposed by browser execution.}
\label{fig:appendix_strategy_risk}
\end{figure}

\autoref{fig:appendix_strategy_risk} is a fixed policy diagnostic. If repair were generic self refinement, one action family should dominate return, reliability, and coverage at the same time. Panel A shows a different structure. Interaction repair gives the strongest local correction when the executed trace exposes a broken control, stalled transition, or failed playability path, but it covers a narrower part of the repairable population. Rewrite covers many structurally weak pages and contributes large recovery, but it is not the most reliable family once the page already has usable structure. Other targeted fixes absorb a broad set of local defects, yet their average lift is smaller. This is why the controller is useful: the objectives do not collapse into one ranking.

Panels B and C expose the same dependency at the individual strategy level. The highest value actions are not mainly visual polish operations; they are interaction, playability, and holistic rewrite actions whose usefulness depends on what the browser trace revealed. This matters for the main claim. A screenshot only judge would see the surface state but would not reliably distinguish a missing click response from a broken game loop or from a page that needs replacement. The state aware controller uses the executed trace to choose among these repair families: replace when there is no stable local target, diagnose when partial structure remains, and preserve when further broad editing is more likely to damage than recover the page.

\subsection{Convergence and Stopping}

\autoref{fig:repair_value_curve} in the main analysis shows that more rounds alone do not explain the improvement. The population moves sharply once a useful repair path is found, and the current candidate can regress after a better checkpoint has already appeared. This is why \method runs every candidate again and exports the best cleared checkpoint, not the final candidate in the loop.

\begin{figure}[H]
\centering
\includegraphics[width=0.96\textwidth]{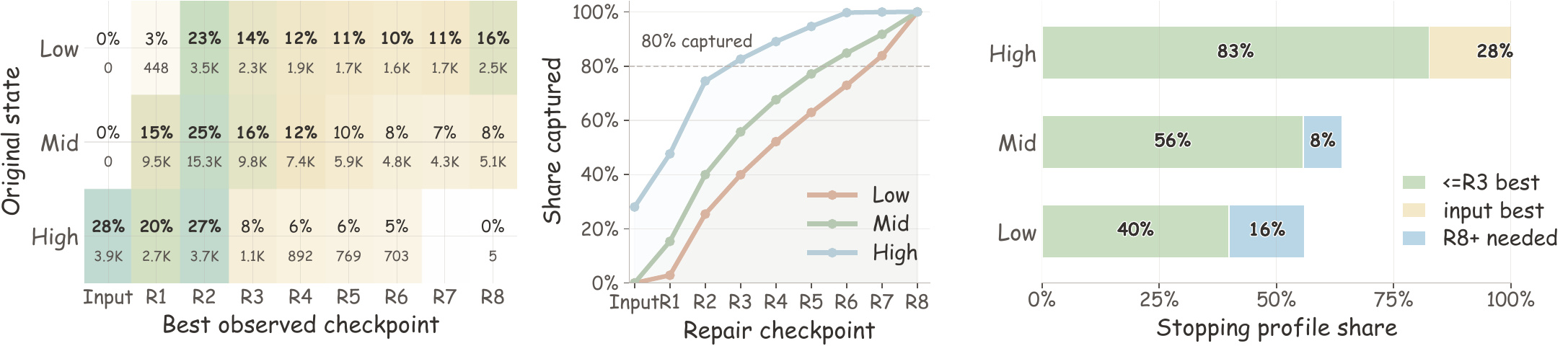}
\caption{State dependent stopping behavior. Panel A locates the first best checkpoint for each original state. Panel B shows how quickly each state captures its best candidate. Panel C summarizes early peaks, input best cases, and late recoveries, motivating preservation for High pages and longer search for weaker pages.}
\label{fig:appendix_best_stop}
\end{figure}

\autoref{fig:appendix_best_stop} explains why stopping must also depend on state. Low pages often need several rounds because the first successful action must reconstruct missing structure before local defects become visible. Mid pages reach their best checkpoint earlier, which matches the diagnosis repair route in the main policy. High pages behave differently: many are already best at input or peak in the first few rounds. Continuing to edit them creates little additional value and mainly opens a regression channel. The figure therefore supports preservation as an active policy choice, not as a failure to improve high scoring pages.

Taken together with \autoref{fig:repair_value_curve}, the appendix figures give the mechanism level evidence behind the corpus and SFT results. The benefit of \method is not that it edits more. It edits after experiencing the page, uses that experience to choose the repair family, verifies each candidate through a fresh execution pass, and stops or preserves when the evidence indicates that further edits are more likely to revisit or damage an already usable page. These controls make repair suitable for data construction: the refined pool grows because weak and partial pages are recovered under a measurable quality gate, while strong pages are protected from unnecessary rewriting.

\subsection{Representative Case Studies}

\begin{figure}[H]
\centering
\includegraphics[width=0.98\textwidth]{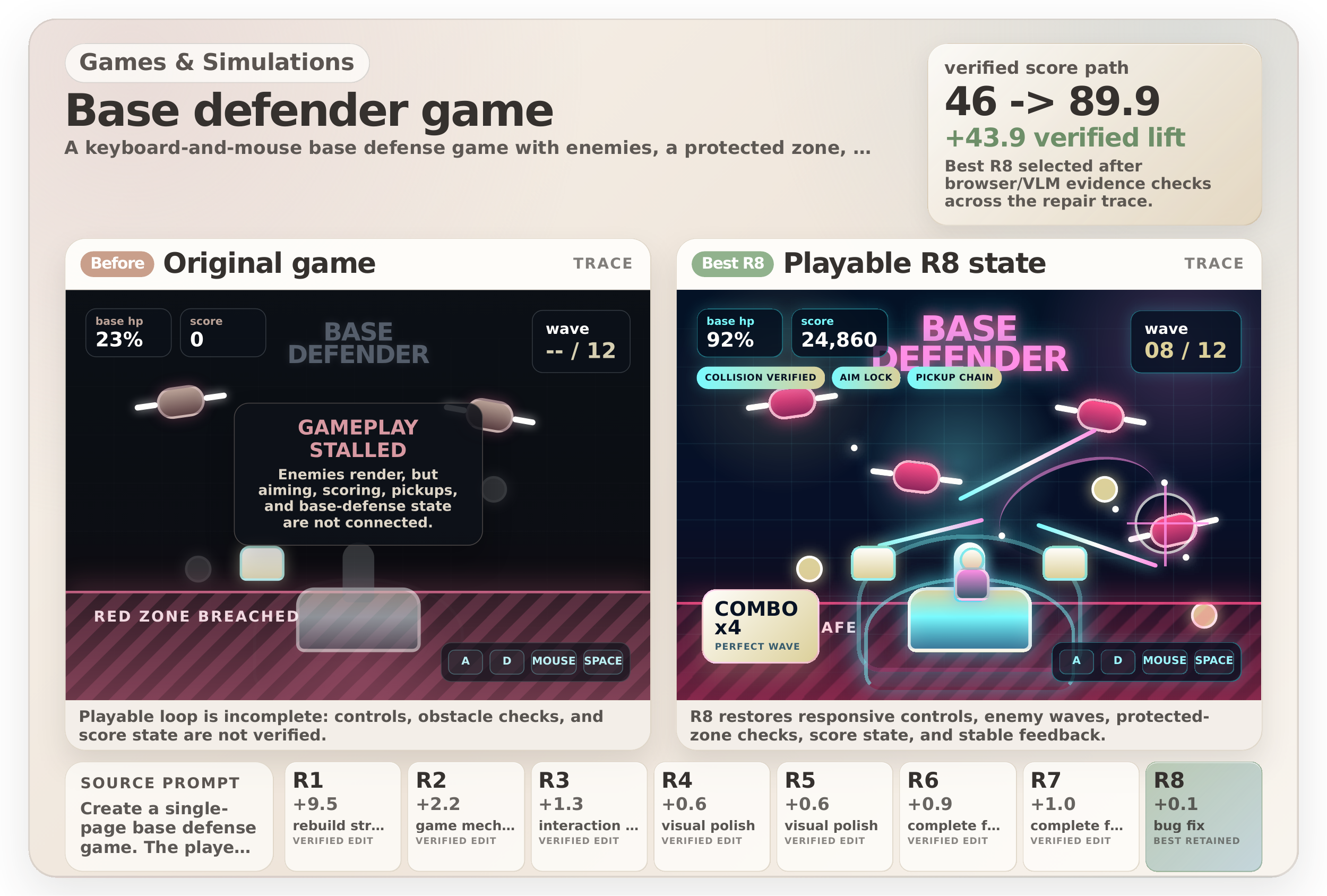}
\caption{Real repair trace for Games \& Simulations. The card contrasts the original broken gameplay state with a playable retained checkpoint and shows the verified repair actions that recover the game loop.}
\label{fig:case_games}
\end{figure}

\begin{figure}[H]
\centering
\includegraphics[width=0.98\textwidth]{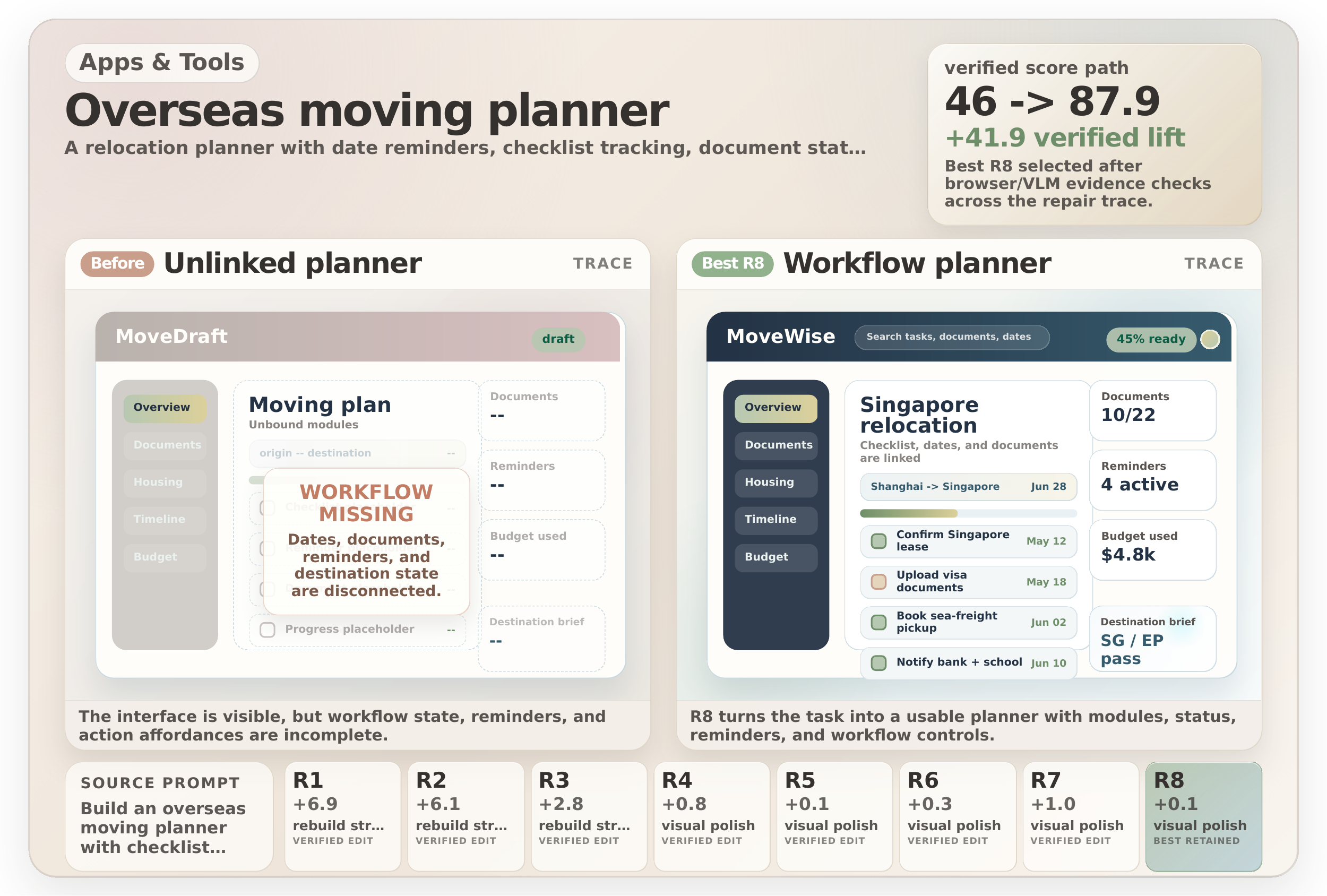}
\caption{Real repair trace for Apps \& Tools. The card shows how an early off target application state is repaired into a usable planner with task state and workflow affordances.}
\label{fig:case_apps}
\end{figure}

\begin{figure}[H]
\centering
\includegraphics[width=0.98\textwidth]{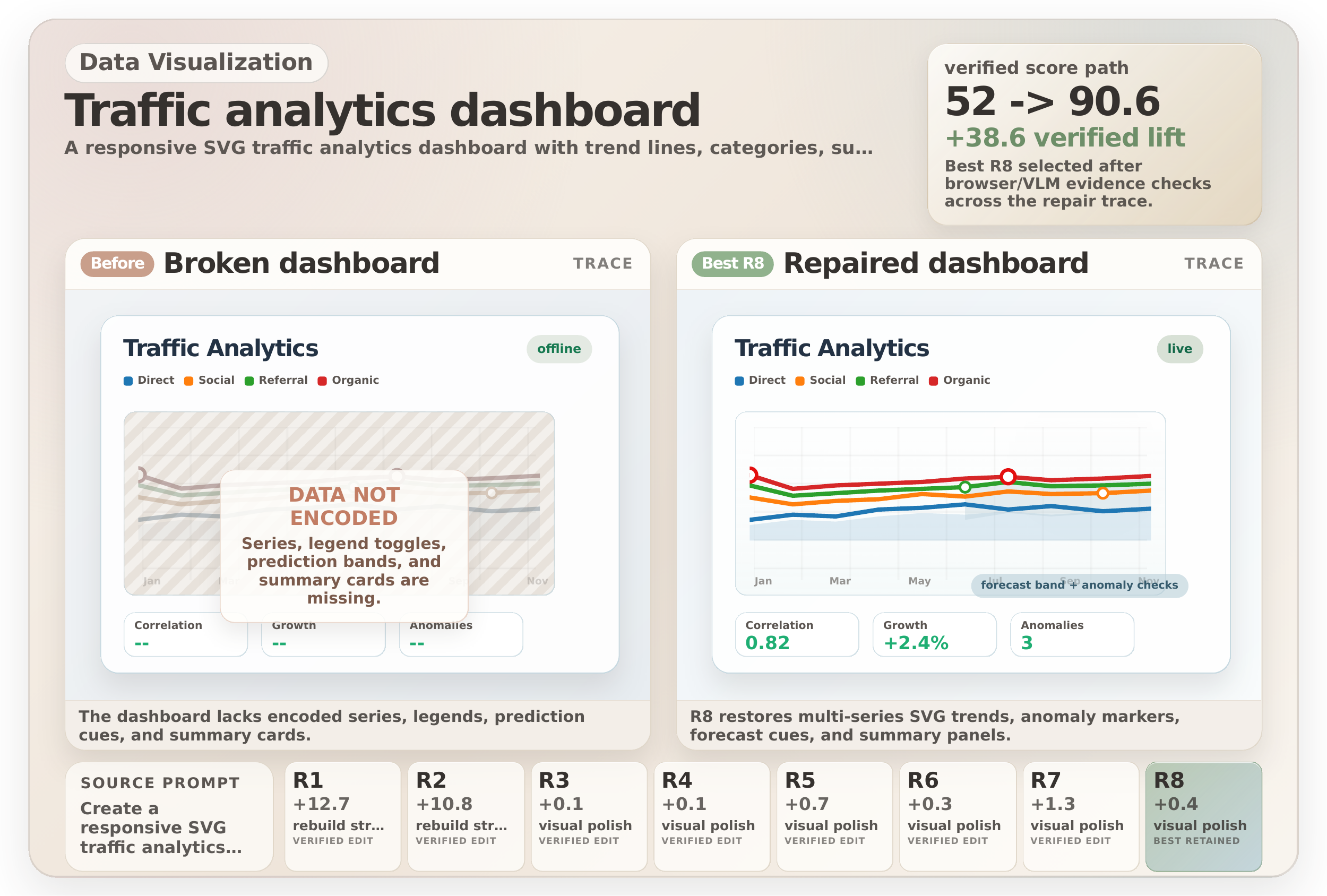}
\caption{Real repair trace for Data Visualization. The card contrasts a missing analytic structure with a retained checkpoint that restores encoded series, legend controls, and summary panels.}
\label{fig:case_data}
\end{figure}

\begin{figure}[H]
\centering
\includegraphics[width=0.98\textwidth]{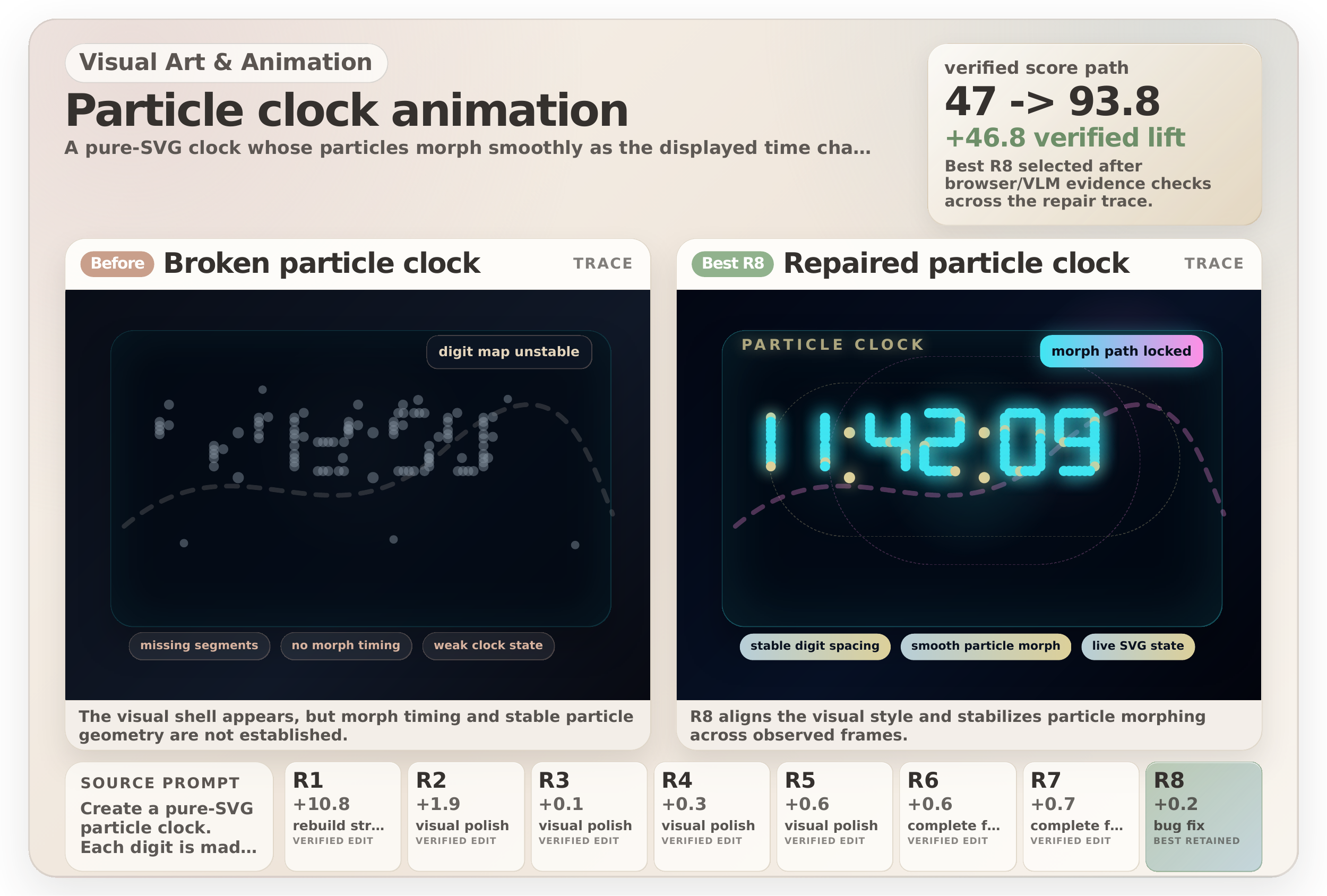}
\caption{Real repair trace for Visual Art \& Animation. The card shows how browser evidence guides the repair from off target motion toward a coherent particle clock animation.}
\label{fig:case_visual}
\end{figure}

\begin{figure}[H]
\centering
\includegraphics[width=0.98\textwidth]{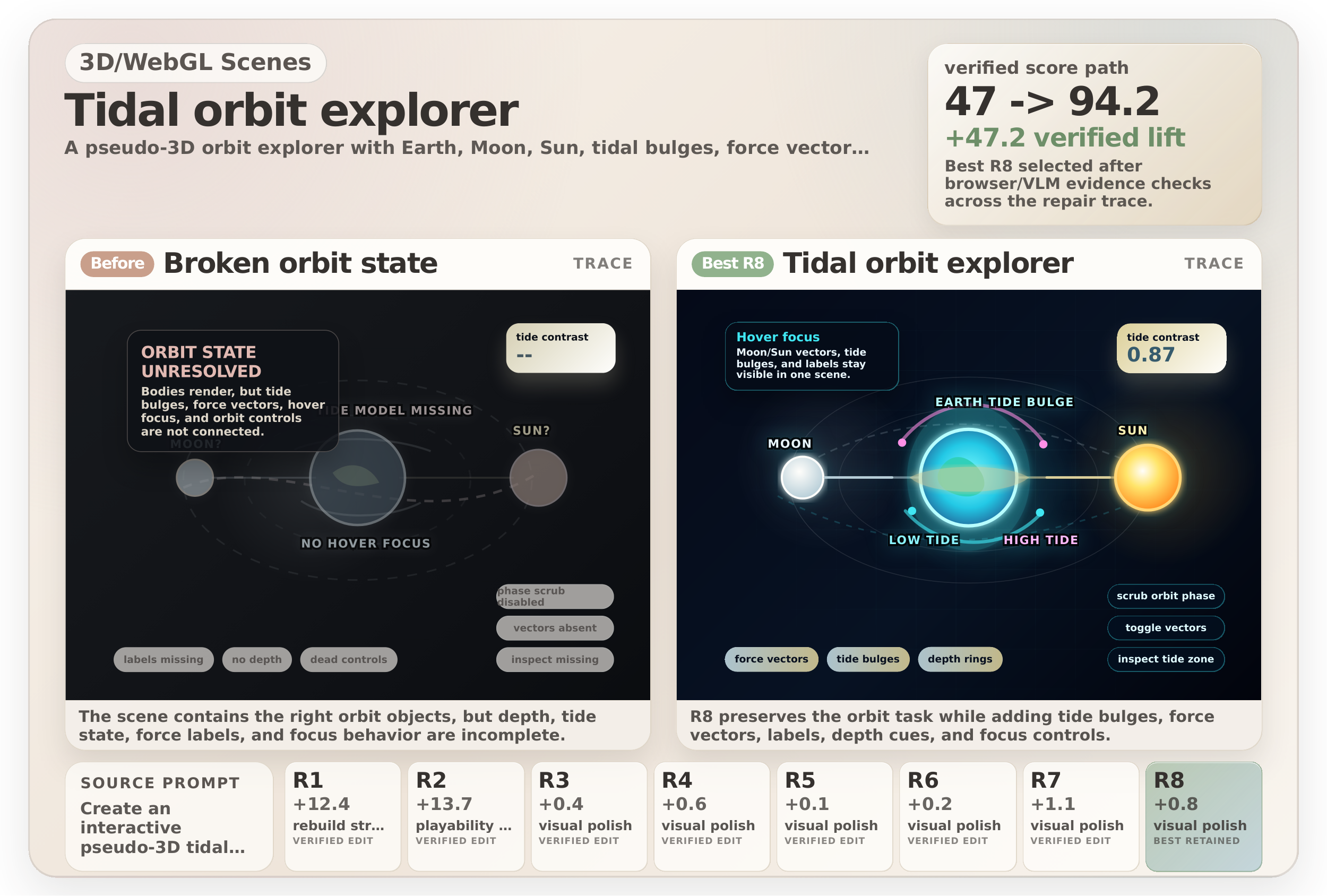}
\caption{Real repair trace for 3D/WebGL Scenes. The card keeps the radial map task and shows how repair adds neon regions, depth rings, labels, and interaction ready SVG structure.}
\label{fig:case_3d}
\end{figure}

\begin{figure}[H]
\centering
\includegraphics[width=0.98\textwidth]{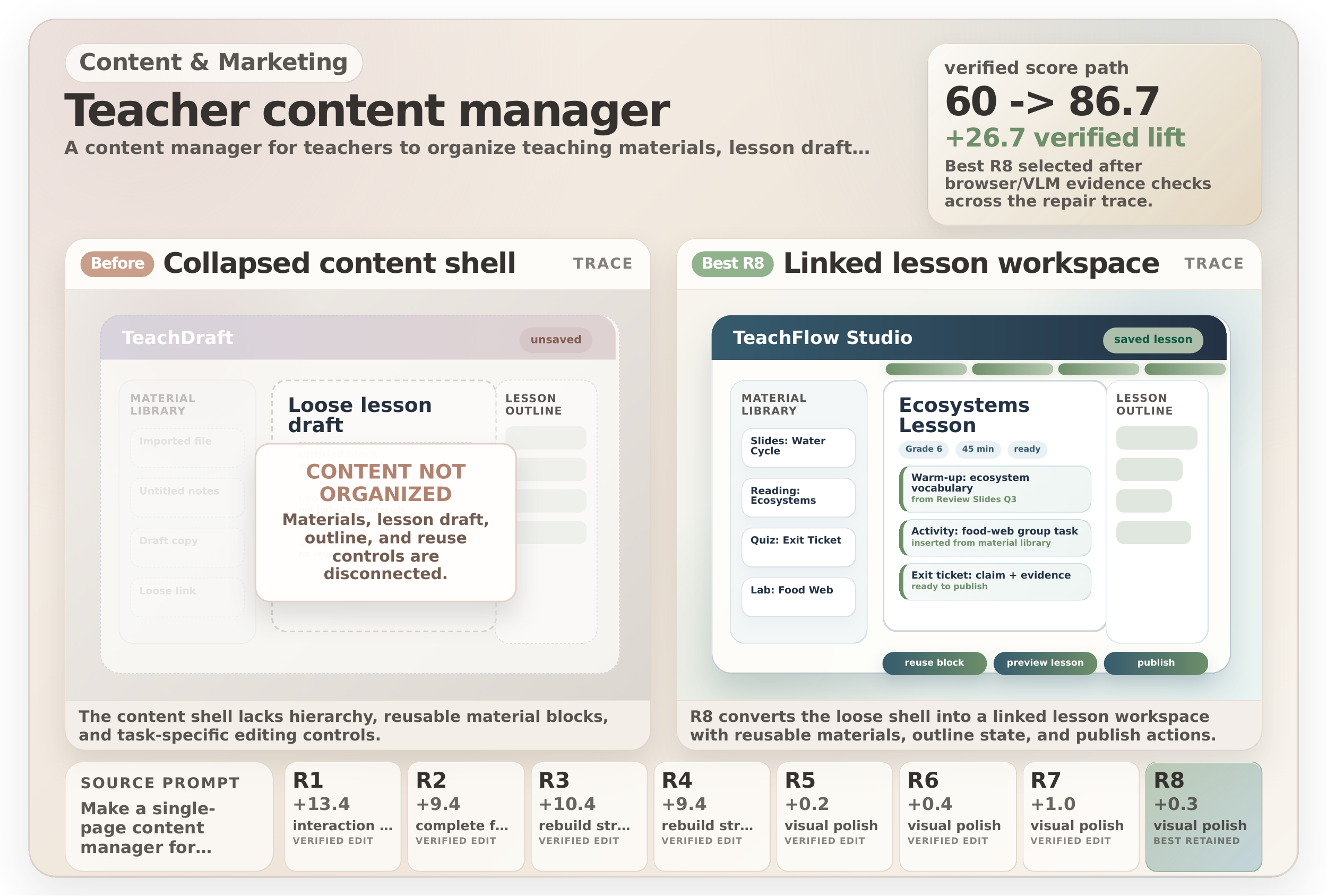}
\caption{Real repair trace for Content \& Marketing. The card shows how repair turns a coarse content shell into a clearer lesson building workspace with reusable materials and scanable hierarchy.}
\label{fig:case_content}
\end{figure}

Figures~\ref{fig:case_games}--\ref{fig:case_content} follow the same semantic family order as \autoref{tab:bench_categories} and show case cards rendered in a browser from selected repair traces. These are qualitative examples rather than additional aggregate evidence; the measured claims come from the trace analyses above. Each card places the original render beside the retained checkpoint, reports the verified score path, and lists the repair action used at every round. The point is to expose what the controller is optimizing: not just a better screenshot, but executable structure, interaction feedback, rendering stability, and visual coherence over a page trajectory.

\FloatBarrier

\section{Additional Details}

This appendix records the benchmark specification, reproducibility boundary, and compact prompt templates used by the evaluation and repair system.

\subsection{SFT Data Construction and Training Routes}
\label{app:sft_data}

The SFT data are derived from 97{,}115 HTML generation prompts collected from browser crawled web tasks and mapped to the same six semantic families used by \benchmark. Kimi-K2.5 generates one HTML page per prompt. Format extraction and browser initialization produce 92{,}418 valid Raw rows with evaluator scores. Of these, 91{,}484 also have completed repair traces; the remaining 934 valid Raw rows are retained only for Raw SFT because they do not have a completed repair trajectory.

The repair controller assigns each traced page to the Low, Mid, or High state used throughout the paper. Low pages usually require structural reconstruction, Mid pages have enough usable structure for diagnosis guided repair, and High pages are primarily preservation cases. A repaired page enters the quality cleared candidate pool only after the retained version is executed again and passes the final gate. This process yields the 63{,}703 page refined candidate pool in Table~\ref{tab:data_funnel}; the final Refined SFT route uses a 40K page export sampled from that pool.

The natural size comparison uses the amount of data each route actually provides. Raw uses the valid generation pool, Filtered uses the original High subset, and Refined uses the 40K repair cleared export. The equal size ablation in Table~\ref{tab:ablation_results} fixes the count at 12{,}392 examples per route, which separates data quality from sample count. All route comparisons keep the prompt response template fixed, so a training example always contains the same user prompt field and one complete HTML response.

All SFT runs use LlamaFactory with full parameter SFT of Qwen3.5 checkpoints. We use the \texttt{qwen3\_5\_text} template, packing, maximum sequence length 32{,}768, cosine learning rate schedule, warmup ratio 0.05, and 3 epochs. The default learning rates are scale specific: $1{\times}10^{-5}$ for 4B, $8{\times}10^{-6}$ for 9B, and $5{\times}10^{-6}$ for 27B. The 4B and 9B routes use 32 GPUs, while the 27B route uses 128 GPUs with ZeRO style full parameter training configuration. These settings are held fixed within each scale; Raw, Filtered, and Refined differ only in their supervision source.

\subsection{Data Provenance, Licensing, and Compute}
\label{app:data_compute}

The 97{,}115 generation prompts are browser crawled task descriptions collected from publicly reachable web pages under an allowlist collection policy. The crawler keeps the task request and task family metadata rather than copying source page assets, account gated content, credentials, payment flows, private services, or personal data. Prompts that require real login, payment, private backend access, or collection disallowed by the source policy are excluded before generation. Kimi-K2.5 is then used to generate one synthetic single file HTML response for each accepted prompt. These provider generated responses are used as synthetic supervision for Qwen SFT under the project provider terms; redistribution is handled separately from internal training. The public artifact therefore releases benchmark files, aggregate trace tables, SFT manifests, sampling scripts, and reconstruction code. Release of the full generated or refined HTML corpus is governed by the corresponding source and provider redistribution terms.

Compute reporting is split by stage in \autoref{tab:pipeline_compute}. Initial generation and repair are API based; request logs retain provider and model identifiers, retry status, timestamps, and token usage when available, but API keys, private endpoints, account identifiers, and billing rates are not released. Local model evaluation uses NVIDIA H20 GPU workers for vLLM serving and a Chromium based browser runner for executable tests. \benchmark evaluation of local SFT checkpoints uses tensor parallel size 8, maximum model length 32{,}768, temperature 0, generation concurrency 4, timeout 600 seconds, and maximum generation length 30{,}000 tokens; browser tests run in full mode over the frozen 400 item benchmark. MiniAppBench evaluation uses Chromium through Playwright on the released validation split with parallel worker count 3 and the same maximum generation length. \autoref{tab:sft_compute} reports the SFT wall clock and H20 card hour accounting from archived LlamaFactory logs.

\begin{table}[H]
\centering
\scriptsize
\setlength{\tabcolsep}{3.2pt}
\renewcommand{\arraystretch}{1.05}
\caption{Pipeline compute and reporting. Local GPU stages use NVIDIA H20 cards; API credentials and private billing metadata are excluded from the release.}
\label{tab:pipeline_compute}
\begin{tabular}{>{\raggedright\arraybackslash}p{0.16\linewidth}>{\raggedright\arraybackslash}p{0.23\linewidth}>{\raggedright\arraybackslash}p{0.51\linewidth}}
\toprule
\textbf{Stage} & \textbf{Compute} & \textbf{Reported settings} \\
\midrule
Initial generation & Provider API & One Kimi-K2.5 call per accepted prompt; logs retain model, timestamp, retry status, and token usage when available. \\
Repair & Provider API + browser runner & Up to 8 rounds and at most 2 candidates per round; every candidate is re-executed before acceptance; logs retain state band, strategy, scores, gate decision, and retained checkpoint. \\
\benchmark eval. & 8 H20 GPUs + browser runner & vLLM tensor parallel 8, max length 32{,}768, temperature 0, generation concurrency 4, timeout 600s, max output 30{,}000 tokens, full 400-item browser evaluation. \\
MiniAppBench eval. & 8 H20 GPUs + Playwright & vLLM tensor parallel 8 for local checkpoints, validation file \texttt{query\_validation\_100.json}, Chromium browser, 3 parallel workers, max output 30{,}000 tokens. \\
SFT & 32 or 128 H20 GPUs & LlamaFactory full parameter SFT, packing, sequence length 32{,}768, 3 epochs, per-device batch 1, gradient accumulation 1, cosine schedule, warmup 0.05. \\
\bottomrule
\end{tabular}
\end{table}

\begin{table}[H]
\centering
\scriptsize
\setlength{\tabcolsep}{4.2pt}
\renewcommand{\arraystretch}{1.02}
\caption{SFT compute from archived training logs. All routes use NVIDIA H20 GPUs; card h is wall clock multiplied by the recorded card count.}
\label{tab:sft_compute}
\begin{tabular}{llrrr}
\toprule
\textbf{Route} & \textbf{Scale} & \textbf{Cards} & \textbf{Wall h} & \textbf{Card h} \\
\midrule
Raw & 4B & 32 & 62.9 & 2{,}014 \\
Filtered & 4B & 32 & 7.5 & 241 \\
Refined & 4B & 32 & 47.4 & 1{,}516 \\
Raw & 9B & 32 & 63.6 & 2{,}035 \\
Filtered & 9B & 32 & 7.7 & 246 \\
Refined & 9B & 32 & 24.4 & 781 \\
Raw & 27B & 128 & 46.6 & 5{,}962 \\
Filtered & 27B & 128 & 5.8 & 740 \\
Refined & 27B & 128 & 18.1 & 2{,}320 \\
\midrule
A1 Raw & 9B & 32 & 8.5 & 272 \\
A2 Filtered & 9B & 32 & 7.6 & 245 \\
A3 Refined & 9B & 32 & 11.7 & 375 \\
A4 Raw & 27B & 128 & 6.3 & 801 \\
A5 Filtered & 27B & 128 & 5.8 & 741 \\
A6 Refined & 27B & 128 & 8.7 & 1{,}110 \\
\bottomrule
\end{tabular}
\end{table}

\subsection{External MiniAppBench Protocol}
\label{app:miniappbench_protocol}

MiniAppBench is used as an external transfer check rather than as part of the repair loop. It is never used for routing, export gating, threshold selection, or checkpoint selection. The full benchmark contains 500 tasks over six domains, but the public test set withholds evaluation references for leaderboard integrity. We therefore report all paper numbers on the released validation file, \texttt{query\_validation\_100.json}, which includes the user queries, difficulty levels, domain labels, and evaluation references needed for local reproduction. The split is fixed before model comparison, external to \benchmark and the SFT corpus, and evaluated with the same MiniAppEval runner for every row in Tables~\ref{tab:main_results} and~\ref{tab:ablation_results}. We use it as evidence of transfer beyond \benchmark-400, not as a replacement for a hidden benchmark submission.

\begin{table}[H]
\centering
\scriptsize
\setlength{\tabcolsep}{5pt}
\renewcommand{\arraystretch}{1.05}
\caption{MiniAppBench external evaluation split. Full counts are from the benchmark release; validation counts are the fixed released split used in this paper.}
\label{tab:miniappbench_protocol}
\begin{tabular}{lrrl}
\toprule
\textbf{Domain} & \textbf{Full} & \textbf{Val.} & \textbf{Description} \\
\midrule
Science & 187 & 31 & Simulators and virtual laboratories \\
Games & 121 & 23 & Puzzles, simulations, and casual games \\
Tools & 57 & 28 & Schedulers, editors, and utilities \\
Visualization & 56 & 4 & Charts and interactive generative art \\
Humanities & 47 & 12 & Skill learning and cultural study \\
Lifestyle & 32 & 2 & Trackers, toys, and roleplay applications \\
\midrule
Total & 500 & 100 & Six-domain external validation split \\
\bottomrule
\end{tabular}
\end{table}

\subsection{\benchmark Specification and Reproducibility}
\label{app:benchmark_spec}

\benchmark is frozen as a benchmark for single file HTML generation that can be executed in a browser. The release contains 400 item identifiers, task family metadata, test case identifiers, test case weights, and browser action sequences. A benchmark item is the atomic public unit shown in \autoref{tab:htmlbench_schema}: models receive only the prompt, while the test cases, weights, and execution traces remain on the benchmark side. The six public families are collapsed from ten source buckets (apps, UI pages, games, data visualization, creative pages, SVG art, 3D/WebGL, content pages, landing pages, and portfolios), which prevents the benchmark from inheriting narrow collection names as evaluation labels.

\begin{table}[H]
\centering
\scriptsize
\setlength{\tabcolsep}{4pt}
\renewcommand{\arraystretch}{1.08}
\caption{\benchmark item and test case schema.}
\label{tab:htmlbench_schema}
\begin{tabular}{ll}
\toprule
\textbf{Field} & \textbf{Purpose} \\
\midrule
\texttt{id} & Stable benchmark item identifier. \\
\texttt{category}, \texttt{sub\_type} & Six family taxonomy plus finer subtype metadata. \\
\texttt{difficulty} & Easy, medium, or hard benchmark stratum. \\
\texttt{prompt} & User facing request for one self contained HTML page. \\
\texttt{has\_interaction} & Whether the task expects active browser behavior. \\
\texttt{test\_cases} & Weighted deterministic browser checks with ordered steps. \\
\texttt{steps} & Actions such as click, hover, key press, resize, JavaScript assertion, and screenshot change checks. \\
\bottomrule
\end{tabular}
\end{table}

The frozen scored pool contains 6{,}000 test cases. The runner executes 16{,}120 steps across those cases, with the largest groups coming from JavaScript assertions, waits, screenshot change checks, screenshots, clicks, text clicks, JavaScript evaluation, resize, key press, and text assertions. These actions are intentionally mixed: some check visible page health, some check semantic DOM state, and others force interaction or responsive layout transitions. Coverage is recorded as metadata and is not a score dimension.

\begin{table}[H]
\centering
\scriptsize
\setlength{\tabcolsep}{4pt}
\renewcommand{\arraystretch}{1.08}
\caption{\benchmark construction and independence controls.}
\label{tab:htmlbench_controls}
\begin{tabular}{p{0.22\linewidth}p{0.70\linewidth}}
\toprule
\textbf{Control} & \textbf{Implementation in the released benchmark} \\
\midrule
Prompt source & Six semantic families with 65 subtypes and balanced easy, medium, and hard strata. The benchmark keeps source bucket metadata for audit, but models are evaluated only from the user facing prompt. \\
Test case rule & Checks must be prompt grounded and observable in the browser: content, controls, state changes, responsive behavior, rendering health, and semantic structure. Tests must not depend on class names, variable names, frameworks, hidden DOM layout, real credentials, payments, or private services. \\
Frozen scoring pool & The scored pool is a fixed 6{,}000 test selection. Runtime scoring depends on the current page execution and weighted test case pass rate, not on model identity or repair history. \\
Quality control & The public files pass the strict benchmark validator with 400 items and no schema or logic errors. The benchmark quality audit reports no duplicate evidence, shallow visual, or cross template risk types in the released files. \\
Train/test separation & An exact normalized prompt match audit between the 400 benchmark prompts and the 91{,}484 traced repair records used for the SFT data analysis finds zero overlaps. \\
\bottomrule
\end{tabular}
\end{table}

The 400 item size is a deliberate benchmark design choice rather than a claim that the open web is exhausted by 400 prompts. Each item is expensive because it carries multiple tests that execute in a browser and because full evaluation includes rendering, probes, keyframes, and optional visual judgment. The interaction heavy mix is also deliberate: static pages can be judged reasonably well by screenshot and layout checks, while the target failure mode of this paper appears when generated pages must respond to user actions. Noninteractive prompts remain in the benchmark and use the scoring profile in \autoref{tab:scoring_dims}, where the interaction budget is not counted against tasks that do not ask for controls.

\begin{table}[H]
\centering
\scriptsize
\setlength{\tabcolsep}{4pt}
\renewcommand{\arraystretch}{1.08}
\caption{\benchmark evaluation protocol.}
\label{tab:htmlbench_eval_protocol}
\begin{tabular}{ll}
\toprule
\textbf{Stage} & \textbf{Benchmark side operation} \\
\midrule
Generation & Model returns one complete HTML document from the prompt. \\
Extraction & Runner extracts valid HTML and records malformed or truncated outputs. \\
Static pass & HTML/CSS/JS hygiene, metadata, dependency, and semantic structure checks. \\
Browser pass & Render, console state, interaction probes, viewport changes, and keyframes. \\
Frozen TCs & Weighted deterministic test cases compute the functionality component. \\
Visual pass & VLM judges visual design from curated executed keyframes in full mode. \\
Aggregation & Dimensions are summed directly to a 100 point score; no normalization after scoring. \\
\bottomrule
\end{tabular}
\end{table}

The visual component is controlled separately from the deterministic browser tests and is not the only basis for interpreting results. In full mode, the Analyst sees screenshots and browser evidence but cannot assign scores. The Scorer receives no screenshots and must score from the Analyst JSON plus objective metrics. The VLM call uses deterministic decoding, and the final composite scorer applies deterministic guardrails before scaling visual design to its 20 point budget. Fast mode skips the VLM phase for low cost regression checks; leaderboard results use full mode. This keeps the visual signal available for design quality while grounding the main interpretation in test case pass rate, component scores, same backbone route comparisons, and MiniAppBench.

Reproducibility is organized around two artifacts. The benchmark artifact consists of the item JSONL files, the frozen test case selection file, the browser runner, the scoring code, validation commands, and configuration templates. The paper artifact additionally reports the aggregate trace tables used for the figures and SFT comparisons, including the state bands, repair outcomes, and iteration level summaries. Together, these files make the benchmark execution path, the repair funnel, and the route comparisons inspectable. Open weight model runs can be reproduced from the released prompts, runner, configs, and independently generated responses, while closed API reference rows should be interpreted with the provider version used at evaluation time.

\subsection{Repair Controller Implementation Details}
\label{app:repair_controller_details}

The repair controller is intentionally a fixed rule system in this paper. It is not trained on \benchmark, and it does not query a learned policy to choose actions. Its input is the structured diagnosis $D(h)$ described in \autoref{tab:repair_diagnosis_schema}; this record is built from browser execution, deterministic tests, visual analysis, and repair history before the LLM receives a prompt.

\begin{table}[H]
\centering
\scriptsize
\setlength{\tabcolsep}{4pt}
\renewcommand{\arraystretch}{1.08}
\caption{Structured diagnosis used by the repair controller.}
\label{tab:repair_diagnosis_schema}
\begin{tabular}{p{0.23\linewidth}p{0.69\linewidth}}
\toprule
\textbf{Field group} & \textbf{Controller evidence} \\
\midrule
Score state & Total score, five component scores, Low/Mid/High band, and distance to the export target. \\
Test failures & Failed test case identifiers, assertion messages, missing requirement items, and pass rate changes from previous rounds. \\
Runtime state & Render success, console errors, JavaScript exceptions, overlays, FPS or frozen motion evidence, and responsive layout failures. \\
Interaction state & Responsive buttons, keyboard bindings, form submission behavior, hover response, drag response, gameplay state changes, and latency. \\
Visual state & Analyst observations, distinctive or generic design signals, mobile and desktop keyframe notes, and contrastive before/after feedback. \\
History & Previous strategies, accepted and rejected candidates, best verified checkpoint, regressions, and preservation list. \\
\bottomrule
\end{tabular}
\end{table}

The operating thresholds have specific roles. Scores below 40 are treated as Low because the implementation often lacks a stable local target; scores from 40 to 79 are treated as Mid because enough structure exists for diagnosis guided repair; scores at or above 80 are treated as High and routed through conservative refinement. The target 97 is an early stop condition for pages that are already strong enough for export. The controller can request at most two candidates per round and always compares them with the current checkpoint. This design prevents uncontrolled edit sampling, keeps regression checks cheap enough to run at corpus scale, and makes the exported page the best checkpoint verified by execution rather than the last generated candidate.

\subsection{Prompt Templates}
\label{app:prompts}

This appendix provides the complete fixed prompt templates used by the benchmark generation, evaluation, browser testing, and repair agents. Braced fields are populated at runtime with task specific evidence, generated HTML, keyframe annotations, probe summaries, previous iterations, or output mode settings; the surrounding instructions, output contracts, and state dependent control logic are shown in full.

\subsubsection{Benchmark Generation}
The generation prompt is intentionally minimal so that \benchmark measures first pass HTML generation under a uniform interface. It enforces a single self contained HTML artifact and forbids explanations or markdown wrappers.

\paragraph{Prompt template.}
For benchmark model evaluation, each model receives the original task as the user message and a minimal system instruction that forces a single self contained HTML response. The generator rejects empty, truncated, or non HTML outputs and retries safety sensitive tasks with a UI only demo suffix when required by the provider.

\begin{promptbox}{Benchmark HTML Generation System Prompt}
You generate benchmark solutions as a single self-contained HTML document.
Return only the final HTML.
Do not include analysis, explanations, markdown fences, or any text before/after
the HTML.
Start with <!DOCTYPE html> when applicable and ensure the document is complete
and closed.

Optional retry suffix for provider-safety blocks:
Build a harmless, front-end-only demo. Use fake placeholder data only. Do not
process, transmit, store, or validate real credentials or payment data; if
payment fields are requested, implement UI-only demo formatting/validation with
synthetic examples.
\end{promptbox}

\subsubsection{Experience Evaluation Agents}
The evaluation stage separates visual observation from scoring. The Analyst receives screenshots and browser evidence but does not assign scores. The Scorer receives no screenshots and must justify each dimension from structured evidence.

\paragraph{Experience Analyst prompt template.}
The Analyst is the only evaluation agent that sees screenshots. It receives the task, static code facts, render and console evidence, interaction probes, keyframe annotations, responsive views, DOM inventory, visible text, and selected screenshots. It returns both visual observations and a requirement checklist; it does not assign scores.

\begin{promptbox}{Experience Analyst Prompt}
You are a meticulous HTML page analyst. You will describe EXACTLY what you see in
the screenshots, report objective facts from probe data, AND verify each task
requirement. Do NOT score. Only observe and audit.

## Task Description
{query}

## Static Analysis
- HTML size: {html_size}; Canvas: {has_canvas}; JS: {has_script};
  CSS: {has_style}; SVG: {has_svg}; rAF: {has_raf}
- Input types detected: {input_types}
- External resources: {ext_list}
- Static issues: {static_issues}

## Render and Runtime Evidence
- Rendered: {rendered}; Title: {page_title}
- Serious console errors: {console_list}
- JS exceptions: {page_err_list}

## Browser Interaction Evidence
- Agent phase: {agent_ran}; steps: {agent_steps}; actions: {agent_actions}
- Confirmed responsive keys: {discovered_keys}
- Keyboard probe: {keyboard_probed}; responsive: {keyboard_responsive}
- Button census: {buttons_responsive}/{buttons_tested} responsive
- Form, drag, gameplay, canvas, animation, audio, and latency evidence:
  {dynamic_experience_fields}

## Screenshots
Each image is a keyframe from automated testing. Review them IN ORDER.
{frame_annotations}

Compare consecutive frames carefully:
- Identical frames after interaction = interaction BROKEN
- Progressive change = interaction WORKING

## Your Task
Part 1: Observation.
Describe the page factually. Cite specific evidence such as frame deltas, latency,
button response rate, keyboard response, canvas fill, and console state. List
distinct visual elements, generic/template-like signals, distinctive design
signals, working features, broken features, interaction quality, and layout notes.

Part 2: Requirement Audit.
Break the task description into prompt-grounded requirements. For each requirement,
return status = done, broken, or missing with evidence. Ignore packaging constraints
such as "single HTML file" unless they cause a user-facing failure. Do not count
optional extras as requirements.

Reply ONLY with valid JSON:
{
  "page_type": "...",
  "visual_state": "...",
  "visual_elements": ["..."],
  "template_like_signals": ["..."],
  "distinctive_design_signals": ["..."],
  "design_specificity": "...",
  "working": ["..."],
  "broken": ["..."],
  "interaction_quality": "...",
  "layout_notes": "...",
  "requirements": [
    {"requirement": "...", "status": "done|broken|missing", "evidence": "..."}
  ],
  "summary": {"total": 0, "done": 0, "broken": 0, "missing": 0}
}
\end{promptbox}

\paragraph{Evidence only scorer prompt template.}
The Scorer receives no screenshots. It reads the Analyst JSON and objective metrics, then assigns the five component score. This separates visual perception from scoring and lets the implementation clamp scores when objective probes contradict the model's text report.

\begin{promptbox}{Evidence-Only Scorer Prompt}
You are a strict HTML quality scorer. You receive structured evidence from a
prior analysis stage plus objective metrics. You have NO screenshots. Score based
ONLY on the evidence provided.

## Task Description
{query}

## Analyst Report
{observer_report_json}

## Requirement Checklist
{task_auditor_report_json}

## Objective Metrics Summary
- Keyboard: probed={keyboard_probed}; responsive={keyboard_responsive};
  keys={keys_responded}
- Button response rate: {button_response_rate_str}
- Canvas: type={canvas_type}; content={canvas_has_content}; fill={canvas_fill_ratio}
- Animation: detected={animation_detected}; fps={fps_quality};
  frame changes={frame_change_rate}
- Latency: avg={avg_latency}; max={max_latency}; timed out={interactions_timed_out}
- Console and JS exceptions: {console_count}; {page_err_count}
- Form, drag, gameplay, structural, and agent evidence: {objective_metric_fields}

## Scoring Rules
1. Evidence only. If a feature is not supported by the Analyst report or probe
   data, do not assume it exists.
2. Functionality is driven by the requirement checklist. Done items raise the
   score; broken items penalize more than missing items because they indicate a
   failed implementation.
3. Interaction is constrained by objective probes. Unresponsive buttons, keyboard
   failure, gameplay with no state change, or timed-out actions cap interaction.
4. Visual design is judged from the Analyst's visual evidence. Polished but
   reusable templates should not receive top visual scores without prompt-specific
   design signals.
5. Code quality is independent of visual quality and depends on runtime cleanliness,
   maintainability, event wiring, and implementation organization.
6. Score conservatively and cite evidence in every reason.

Return JSON with the five dimensions and total:
{
  "rendering": {"score": 0, "reason": "..."},
  "visual_design": {"score": 0, "reason": "..."},
  "functionality": {"score": 0, "reason": "..."},
  "interaction": {"score": 0, "reason": "..."},
  "code_quality": {"score": 0, "reason": "..."},
  "total_score": 0,
  "bugs": ["specific observed bug"],
  "missing_features": ["missing prompt-grounded feature"],
  "highlights": ["working feature to preserve"],
  "improvement_hints": ["actionable repair hint"],
  "summary": "..."
}
\end{promptbox}

\paragraph{Browser use interaction tester prompt template.}
When the optional agent test phase is enabled, the browser agent receives a task specific interaction protocol. The prompt is augmented with detected input modes and, for games, keys that were already verified by the key scan.

\begin{promptbox}{Automated Interaction Tester Prompt}
You are an automated HTML page quality tester. Thoroughly test the page and
produce a detailed report.

Visit this URL: {page_url}

## What this page should do
{query}

## Testing procedure
1. Load the page and observe initial rendering.
2. Activate entry points such as Start, Submit, Play, Enter, or OK.
3. Test every feature mentioned in the task:
   - click buttons and interactive elements
   - fill and submit forms
   - test navigation, menus, tabs, pagination, and modes
   - observe and interact with animated or canvas-based content
   - navigate through multiple states or views
4. Complete at least one full user workflow.
5. Report unresponsive elements, visual glitches, broken layout, on-page errors,
   and missing content.

{interaction_guide}

## Report
1. Rendering: initial page state and visible failures.
2. Feature status: working, partial, broken, or missing.
3. Bug list: what happened versus what should have happened.
4. Missing features.
5. Overall quality: Excellent, Good, Fair, Poor, or Broken.
\end{promptbox}

\subsubsection{State Aware Repair Agents}
The repair stage receives structured browser evidence collected from the current page. The controller selects the strategy, while the prompt exposes the current score, evidence, prior attempts, and preservation constraints to the code generating model.

\paragraph{State aware repair prompt template.}
All repair strategies share the same prompt structure: current state, scores, previous attempts, objective probe evidence, Analyst evidence, requirement checklist, preservation list, and the current HTML. The controller changes the strategy specific instruction block according to page state and diagnosis. Low pages use rewrite oriented prompts, Mid pages use diagnosis guided repair prompts, and High pages use conservative refinement prompts.

\begin{promptbox}{State-Aware Repair Prompt Template}
You are an expert HTML/CSS/JavaScript developer. Improve the current page under
the strategy selected by the repair controller: {strategy_name}.

## Task
{query}

## Current page state
Total score: {score}/100
Dimension scores: rendering={rendering}, visual_design={visual_design},
functionality={functionality}, interaction={interaction}, code_quality={code_quality}

## Previous repair attempts
{prev_iterations}

## Objective probe evidence
{probe_evidence}

## Visual and requirement evidence
{observer_evidence}
{requirement_checklist}
{visual_context}
{contrastive_feedback}

## Problems to fix
{issues_or_missing_features}

## Features that must be preserved
{preservation_list}

## Strategy-specific instruction
{strategy_instruction}

Examples:
- Holistic rewrite: rebuild a complete implementation when the page lacks a
  stable local target, while preserving any verified useful structure.
- Feature completion: implement all broken or missing requirements without
  removing working features.
- Interaction repair: fix event wiring, input handling, latency, feedback, and
  state transitions without changing unrelated visuals.
- Game repair: target the diagnosed layer such as input, game loop, canvas
  rendering, overlay state, or gameplay logic.
- Visual enrichment: improve typography, layout, palette, depth, animation, and
  responsive polish without breaking verified behavior.
- High-state refinement: make small additive patches and avoid broad rewrites.

## Current HTML
```html
{html}
```

## Output rule
If the selected mode is patch mode, return JSON patches:
{"patches": [{"old_str": "exact substring", "new_str": "replacement"}]}

If the selected mode is rewrite mode, return only the complete HTML file.
\end{promptbox}

\paragraph{Probe driven game repair prompt template.}
Game pages receive a more specific prompt when probes isolate the failed layer. This prompt is used only when the controller detects game like structure and objective evidence such as keyboard response, requestAnimationFrame calls, canvas content, overlays, or game state changes.

\begin{promptbox}{Probe-Driven Game Repair Prompt}
You are an expert HTML/CSS/JavaScript game developer. Automated probes identified
the failing layer: {game_layer}. Fix that layer without rewriting unrelated
systems.

## Task
{query}

## Current state
Total score: {score}/100
Probe evidence: {probe_evidence}
Working features to preserve: {preservation_list}

## Root-cause checklist
If {game_layer}=input:
1. canvas is not focusable or lacks tabindex
2. keydown/keyup listeners are attached to the wrong element
3. preventDefault is missing for arrow keys
4. key state is not read inside the game loop

If {game_layer}=loop:
1. requestAnimationFrame is never called
2. loop starts only after a user action
3. update or draw throws and stops the loop

If {game_layer}=canvas:
1. canvas width or height is zero
2. getContext is missing or called before the DOM is ready
3. clearRect runs without redraw
4. overlay or CSS hides the canvas

If {game_layer}=overlay:
1. game-over or modal screen is visible on load
2. start screen cannot be dismissed
3. initial score, lives, or state incorrectly trigger a terminal state

If {game_layer}=gameplay:
1. collision, scoring, state machine, timer, level progression, or physics is wrong
2. input works but game state does not change correctly

## Current HTML
```html
{html}
```

Return the requested patch or full HTML according to {output_mode}.
\end{promptbox}

\paragraph{Contrastive visual feedback prompt template.}
After a candidate is run again, \method can compare before/after keyframes and inject the result into the next repair prompt. This turns the loop from blind retry into visual, state aware correction.

\begin{promptbox}{Contrastive Visual Feedback Prompt}
You are a visual quality analyst comparing two versions of an HTML page. The page
was modified between BEFORE and AFTER. Identify what improved, what regressed,
and what remained unchanged.

## Task
{query}

## Score change
{score_before} -> {score_after} ({delta})
Dimension deltas: {dim_deltas_str}

## Frame pairs
{pair_count} paired screenshots are provided below. For each pair, BEFORE is the
old version and AFTER is the new version at the same interaction state.

Classify each visible difference:
- IMPROVED: broken, missing, or ugly before; fixed, present, or better after
- REGRESSED: working or good before; broken, worse, or missing after
- UNCHANGED: same problem visible in both versions

Focus on functional and visual differences, not minor pixel shifts.

Reply ONLY with JSON:
{
  "improved": ["..."],
  "regressed": ["..."],
  "unchanged_issues": ["..."],
  "priority_fix": "single most important thing to fix next"
}
\end{promptbox}

\paragraph{Visual diagnosis and verification prompt templates.}
For high scoring pages, \method can request a targeted visual diagnosis and then verify whether a visual enrichment candidate improves the page without breaking behavior.

\begin{promptbox}{Visual Diagnosis and Verification Prompts}
Diagnosis:
You are a visual quality expert. Examine this HTML page screenshot and identify
specific visual issues preventing it from reaching professional quality.

Task: {query}
Current score: {score}; visual_design={visual_design}; rendering={rendering}

Focus on color palette, typography, spacing, layout, visual depth, polish,
responsive quality, hover states, transitions, and micro-animations. Do NOT
suggest functionality changes.

Return JSON:
{"issues": ["..."], "suggestions": ["..."], "css_focus_areas": ["..."]}

Verification:
You are a visual quality analyst comparing BEFORE and AFTER versions of an HTML
page modified for visual quality.

Task: {query}
Before score: {score_before}; After score: {score_after}

Compare polish, layout integrity, content completeness, and whether interactive
controls remain visible and properly styled.

Return JSON:
{"improved": true, "functional_regression": false,
 "improvements": ["..."], "regressions": ["..."]}
\end{promptbox}

\end{document}